\documentclass[aps,prl,amsmath,twocolumn,superscriptaddress,letterpaper,floatfix]{revtex4-1}
\usepackage{graphicx,color}
\usepackage{verbatim}
\usepackage{amssymb}   
\usepackage{amsmath}
\usepackage{amsfonts}
\usepackage{mathdots}
\usepackage{hyperref}
\usepackage{epsfig}

\usepackage{braket}
\usepackage{bm}

\begin{document}

\title{Anisotropic Moir\'e Fractional Chern Insulators and Their Phase Transitions}

\author{Bo Peng}
\affiliation{Division of Physics and Applied Physics,
Nanyang Technological University, Singapore 637371}

\author{Jin-Xin Hu}\thanks{jhuphy@ust.hk}
\affiliation{Department of Physics, Hong Kong University of Science and Technology, Clear Water Bay, Hong Kong, China}
\affiliation{Center for Theoretical Condensed Matter Physics, The Hong Kong University of Science and Technology, Clear Water Bay, Hong Kong SAR, China}

\begin{abstract}
Recently, moir\'{e} heterostructures have been observed to host fractional Chern insulator (FCI) phases at zero magnetic field. In this work, we show that the FCI phases are robust against moderate lattice anisotropy, while sufficiently strong anisotropy drives quantum phase transitions from incompressible topological phases to competing charge-density-wave (CDW) and Fermi-liquid (FL) phases. Specifically, in the case of twisted transition metal dichalcogenides (such as MoTe$_2$), the anisotropy arises from an interlayer momentum shift induced by heterostrain. We find that increasing anisotropy suppresses the many-body topological gap, eventually destabilizing the fractional topological phase. Beyond a critical anisotropy value, the system gradually transitions to a symmetry-broken state characterized by stripe-like CDW order. It has a Landau Level counterpart, where effective mass anisotropy provides an additional tuning knob for the fractional quantum Hall state; sufficiently strong anisotropy drives a transition to a charge-ordered phase. Moreover, in the anisotropic ideal Chern band model with lattice stretching, a Fermi liquid phase emerges as anisotropy increases. Our results establish that anisotropy provides a direct route for engineering and exploring competing correlated phases in FCI states based on moir\'{e} materials.

\end{abstract}

\maketitle

\emph{Introduction.}---The interplay between topology and correlation has emerged as a central topic in modern condensed matter physics, driving the search for exotic phases of matter beyond the Landau paradigm~\cite{fradkin2013field}. The fractional quantum Hall effect (FQHE) stands as the paradigmatic example, hosting topologically ordered states with fractional excitations and robust chiral edge modes~\cite{stormer1999fractional,stormer1999nobel,moore1991nonabelions,bolotin2009observation,cage2012quantum}. In recent years, the discovery of fractional Chern insulators (FCIs) in moir\'{e} heterostructures has revolutionized this landscape, demonstrating that the remarkable physics of the FQHE can be realized at zero magnetic field in lattice systems with topological flat bands~\cite{ju2024fractional}. The flat bands are formed by quenching the kinetic energy at certain twist angles, thereby interaction effect becomes dominate~\cite{bistritzer2011moire}. These breakthrough observations in twisted transition metal dichalcogenides~\cite{cai2023signatures,park2023observation,zeng2023thermodynamic,xu2023observation,redekop2024direct} and graphene-based moir\'{e} systems~\cite{lu2024fractional,xie2021fractional,xie2025tunable,aronson2025displacement,waters2025chern} have established a new platform for exploring fractionally charged excitations and anyonic statistics in highly tunable solid-state devices~\cite{regnault2011fractional,wu2012zoology,bergholtz2013topological,grushin2014floquet,wu2012zoology,wu2013bloch,lauchli2013hierarchy,neupert2015fractional,behrmann2016model}.

A fundamental aspect of the FQHE, inherited by its lattice counterparts, is that the properties of the many-body ground state are intimately linked to the geometry and anisotropy of the underlying interactions.~\cite{yang2013geometry}. In conventional two-dimensional electron gases (2DEGs) under magnetic fields, the anisotropic effective mass and Coulomb interactions are known to drive transitions from topological quantum Hall liquids to charge-density-wave (CDW) phases, including Hall-smectic, stripe, bubble, stripe-crystal, Wigner-crystal, and quasi-one-dimensional crystal orders~\cite{zhu2017anisotropy,he2021charge,wang2012fractional}. This competition among collective phases reflects the delicate interplay between the correlation and interaction anisotropy: strong anisotropy drives an instability toward translational-symmetry-breaking stripe or bubble orders, which may equivalently be viewed as arising from an anisotropically distorted composite-fermion Fermi sea~\cite{ippoliti2017numerical}.

\begin{figure}
		\centering
		\includegraphics[width=1\linewidth]{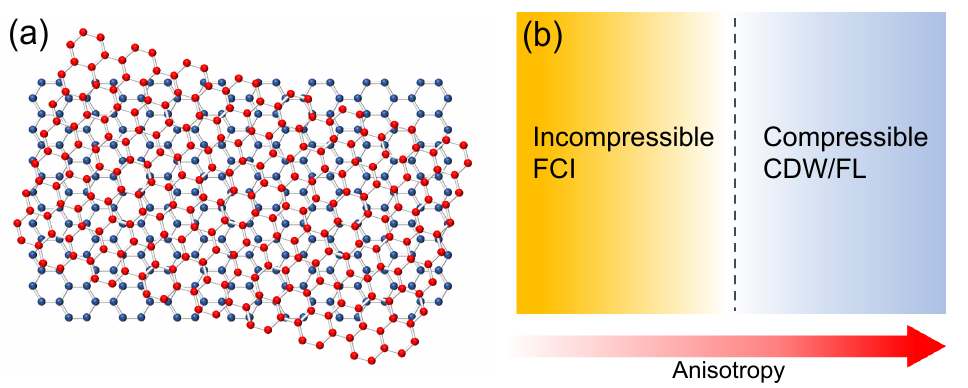}
		\caption{Schematic picture of moir\'{e} heterostructure formed by twisted bilayer honeycomb lattice such as graphene or transition metal dichalcogenides. (b) When anisotropy is introduced, there is the phase transition from incompressible phase (FCI) with topological gap to compressible phase (CDW or Fermi liquid).} 
		\label{fig:fig1}
\end{figure}

Moir\'{e} systems offer a uniquely tunable platform for exploring such anisotropy-driven transitions~\cite{jin2021stripe,xie2025tunable}. 
The exceptional flexibility of van der Waals heterostructures arises from the ability to tune electronic band structure and geometry through twist angle, stacking configuration, and external fields~\cite{carr2017twistronics,wang2021stacking,yu2025engineering}. 
Crucially, anisotropy in moir\'{e} flat bands can be controllably induced by stretching the moir\'{e} superlattice or by interlayer momentum shifts from lattice mismatch or heterostrain~\cite{bi2019designing,hu2022nonlinear}. 
These effects fundamentally alter the quantum geometry of the flat Chern band and the projected interaction, thereby providing a knob to tune between competing ground states. While the incompressible topological phase remains robust against moderate anisotropy, sufficiently strong anisotropy induces increasingly directional correlations and ultimately destabilizes the topological phase. The system may then transition to either a translation-symmetric compressible Fermi liquid or a translation-symmetry-broken CDW phase. 
This behavior mirrors anisotropic FQHE physics~\cite{yang2017generalized,mulligan2010isotropic,wang2012fractional,yang2012band}, now realized in a lattice framework with nontrivial band topology.

In this work, we theoretically investigate the emergence of anisotropic gapped topological phases and their transition to CDW phases or a Fermi liquid under the influence of externally imposed anisotropy (see Fig.~\ref{fig:fig1}). Specifically, we introduce two distinct types of anisotropy: (i) interlayer momentum shifts induced by heterostrain, (ii) geometric anisotropy from stretching the moir\'{e} lattice. Focusing on twisted MoTe$_2$ and chiral twisted bilayer graphene (cTBG) with ideal flat Chern band, we explore the anisotropy-driven phase transitions in topological moir\'{e} bands. Our results reveal that heterostrain serves as a controllable knob to tune between topological and charge-ordered or Fermi liquid phases, with intermediate regimes that may host exotic uncharacterized phases. In addition, the anisotropic topological phases studied in this work share some similarities with the Landau level (LL) behavior, and can also help to understand the underlying mechanisms of the LL system. Our work also demonstrates anisotropic moir\'{e} FCIs as a rich platform for exploring the interplay of topology, correlations, and geometry.

\begin{figure}
\centering
\includegraphics[width=1.0\linewidth]{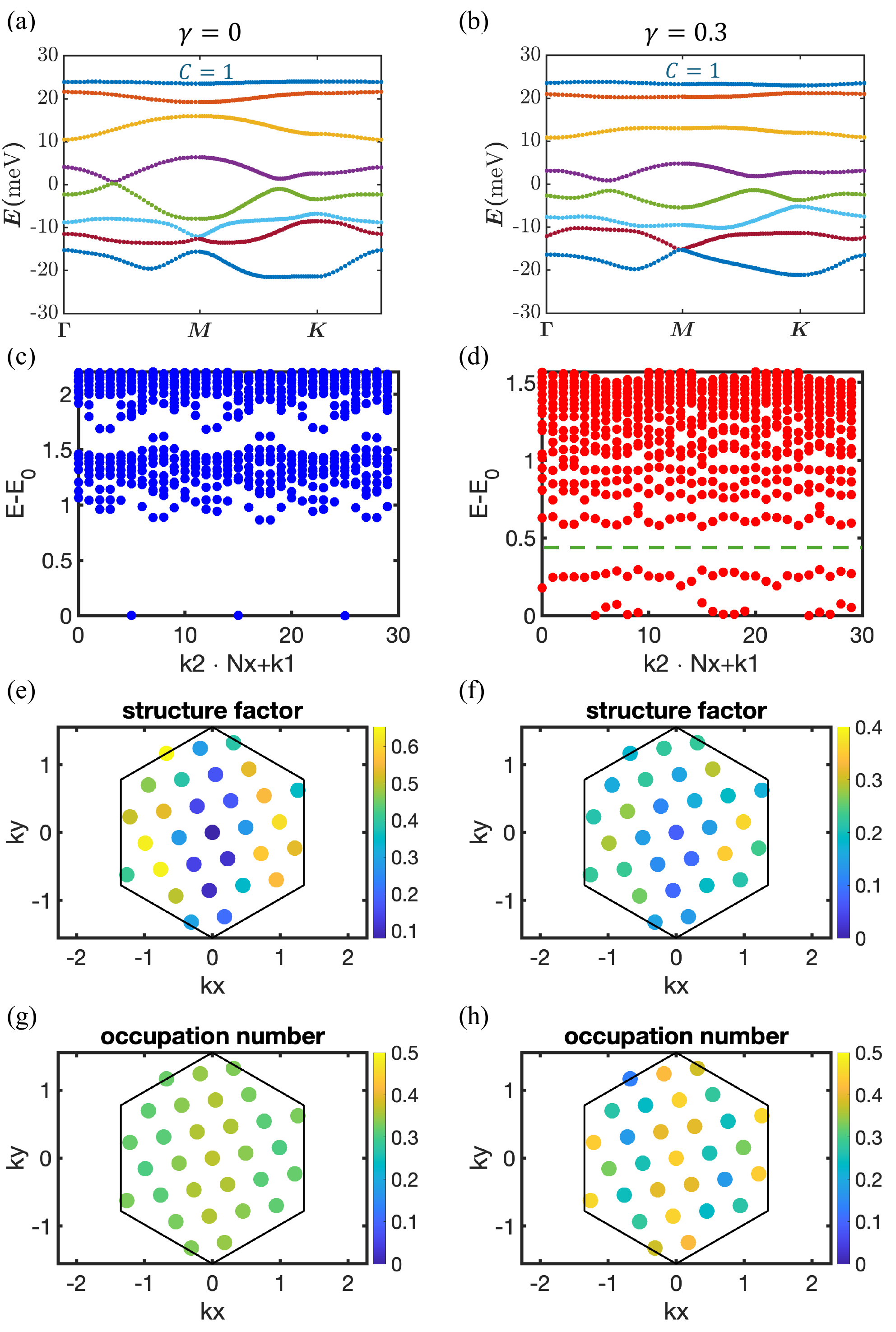}
\caption{ Dispersion of the single-particle Hamiltonian, spectra of the screened Coulomb interaction projected onto the Chern band, Structure factor, and occupation number of ground states. (a,b) Energy dispersion of Eq. (1) without and with internal magnetic field modulation, corresponding to $\gamma = 0$ and $\gamma = 0.3$, respectively. (c,d) Spectra of the screened Coulomb interaction projected onto the Chern band for $(\gamma = 0)$ and $\gamma = 0.3$, respectively. (e,f) Structure factor of the ground states under the projected screened Coulomb interaction for $\gamma = 0$ and $\gamma = 0.3$, respectively. (g,h) Orbital occupation number of the ground state for $\gamma = 0$ and $\gamma = 0.3$, respectively.
}
\label{fig2}
\end{figure}

\emph{Twisted MoTe$_{2}$ under interlayer momentum shift.}---A particular class of bilayer semiconductors based on TMDs has recently garnered significant research attention due to its diverse topological and correlated phases, observed both experimentally and theoretically. Moir\'{e} superlattices formed by twisting a bilayer introduce a long-wavelength periodic structure characterized by the moir\'{e} lattice constant $a_M = a / [2 \sin(\theta/2)]$, where $a$ is the original lattice constant and $\theta$ is the twist angle between the two layers. We shall consider an R-type stacked twisted bilayer. The continuum model Hamiltonian for the K valley reads
\begin{equation}
    H_K=\left( \begin{array}{cc}
        H_b & \Delta_T(\textbf{r}) \\
        \Delta_T^{\dagger}(\textbf{r}) & H_t 
    \end{array}
    \right),
\end{equation}
in which $H_{b(t)}=-\frac{\hbar^2(\textbf{k}-K_{b(t)}+\gamma \bm{q}_{b(t)})^2}{2m^*}+\Delta_{b(t)}$, is the bottom (top) layer Hamiltonian subjected to a moir\'{e} potential. $\Delta_{b(t)}(\textbf{r})=2\nu \sum_{i=1,3,5}cos(\textbf{G}_j\cdot \textbf{r}\pm\psi)$, in which positive for the bottom layer and negative for the top layer.  $\textbf{G}=\frac{4\pi}{\sqrt{3}a_M}\left(cos(\frac{\pi(j-1)}{3}),sin(\frac{\pi(j-1)}{3})\right)$. And $\gamma \bm{q}_{b(t)}$ is an internal magnetic field between two layers of the material, in which $\gamma$ can be tuned from 0 to $0.3$, and $\bm{q}_{b}=-\bm{G}_1$, $\bm{q}_{t}=\bm{G}_1$, where $\bm{G}_1$ is one of the reciprocal lattice vectors of the moir\'{e} superlattice. The interlayer tunneling is dictated by three-fold rotational symmetry as $\Delta_T(\textbf{r})=w(1+e^{-i\textbf{G}_2\cdot \textbf{r}}+e^{-i\textbf{G}_3\cdot \textbf{r}})$. With the parameters $m^*=0.5m_e,\theta=2.8^{\circ}$, $\nu=10\mathrm{meV}$, $\psi=179.2^{\circ}$, and $w=-10\mathrm{meV}$, the dispersion is given in Fig.~\ref{fig2}(a) and (b), with $\gamma=0$ and $\gamma=0.3$, respectively.

Before studying interaction effects within the moir\'{e} flat bands, we provide a more physical interpretation of the interlayer momentum shift. We note that heterostrain in twisted bilayers breaks the $C_3$ rotational symmetry. Strain shifts the Dirac points in opposite directions for the bottom and top layers. The strain tensor is given by
\begin{equation}
  \mathcal{E} =
  \begin{pmatrix}
    \epsilon_{xx} & \epsilon_{xy} \\
    \epsilon_{xy} & \epsilon_{yy}
  \end{pmatrix}.
\end{equation}
Under heterostrain, the reciprocal lattice vectors transform as $\bm{G}_i' = (I - \mathcal{E}^T)\bm{G}_i$, and the momentum-space separation between the two Dirac fermions is $\delta\bm{q} = \frac{\sqrt{3}\beta}{2a_0} (\epsilon_{xx} - \epsilon_{yy},\; -2\epsilon_{xy})$. Assuming a simple diagonal strain tensor $\mathcal{E} = \mathrm{diag}(\epsilon, -\epsilon)$, the magnitude of the interlayer momentum shift becomes $|\delta\bm{q}|/|\bm{G}| = 3\beta \epsilon/(4\pi \theta)$, where $\beta = 2.3$ for typical TMDs~\cite{bi2019designing}. For small $\theta$, the interlayer momentum shift becomes significant in the presence of strain. Another effect induced by strain is the deformation of the reciprocal lattice. We find $|\bm{G}_2'|/|\bm{G}_1'| = 1 + 3\epsilon/2$. For $\epsilon \lesssim 5\%$, the lattice deformation is weak, so we consider only the momentum shift in what follows.

What makes the flat bands so interesting is that they amplify the effect of interactions. When three energy scales obey $\Delta \gg V_{int} \gg BW$, in which $\Delta$ is the energy gap between two bands, $V_{int}$ is the interaction among particles, and $BW$ is the bandwidth of one single band, the interaction dominates within a single band, causing strongly correlated phase. Thus we project a two-body interaction onto the Chern band, 
\begin{equation}
    \hat{V}_{int}=\sum_{\bm{k}_1,\bm{k}_2,\bm{q}} V(\bm{q})c^{\dagger}_{\bm{k}_1} c^{\dagger}_{\bm{k}_2} c_{\bm{k}_1+\bm{q}} c_{\bm{k}_2-\bm{q}},
\label{twobody}
\end{equation}
in which, 
\begin{equation}
    V(\bm{q})= \frac{2\pi e^2\tanh(qd)}{\epsilon_0 \epsilon_r q},
\end{equation}
where $q=|\bm{q}|$ and $d$ is the vertical distance between the
top (bottom) metallic gate and twisted bilayer MoTe$_2$, $\epsilon_0$ is the vacuum permittivity, and $\epsilon_r$ is the relative dielectric constant. 

The spectra computed via exact diagonalization (ED) are shown in Figs.~\ref{fig2}(c) and (d) for $\gamma = 0$ and $\gamma = 0.3$, respectively. In the calculation, we adopt a $5 \times 6$ torus (i.e., $N_x = 5$, $N_y = 6$) with $N_e = 10$ electrons. At $\gamma = 0$, the interlayer momentum shift vanishes. Three quasi-degenerate ground states emerge, sharing the many-body Chern number $C = 1$, indicating a quantized Hall conductivity $\sigma_{xy} = \frac{1}{3} \frac{e^2}{h}$. As the momentum shift gradually increases, the anisotropic FCI phase persists, manifested by the conservation of Hall conductivity. However, as the anisotropy is further increased to $\gamma=0.3$,, the analogous Laughlin phase destabilizes, as evidenced by the closing of the topological gap. In this case, it is interesting to see that, unlike the well-known Laughlin states on the torus, the counting of ground states depends on the system size and aspect ratio. For example, for a $5\times 6$ and $N_e=10$ system, the counting of the CDW ground states is 45.  This implies that this phase is a CDW phase (more details, see Supplementary Materials). 

The structure factor is the Fourier transform of the spatial density distribution, defined as
\begin{equation}
    S(\bm{q})=\frac{1}{N_e}\bigg[\bra{\Psi_0}\rho_{-\bm{q}}\rho_{\bm{q}}\ket{\Psi_0}-\bra{\Psi_0}\rho_{-\bm{q}}\ket{\Psi_0}\bra{\Psi_0}\rho_{\bm{q}}\ket{\Psi_0}\bigg],
\end{equation}
where $\ket{\Psi_0}$ is the ground state of the system. As shown in Fig.~\ref{fig2}(e), at $\gamma = 0$ (no interlayer momentum shift), the structure factor of the screened Coulomb interaction projected onto the Chern band exhibits a ring-like peak, indicating the presence of a topological phase. In contrast, at $\gamma = 0.3$ (Fig.~\ref{fig2}(f)), the structure factor peaks become sharply discretized, indicating a CDW phase. As strong evidence for the phase transition between the FCI and CDW phases, the Bloch state occupation numbers $n_k$ are shown in Figs.~\ref{fig2}(g) and (h). At $\gamma = 0$ (Fig.~\ref{fig2}(g)), the electrons are evenly distributed across momentum space, verifying the emergence of an incompressible topological phase. At $\gamma = 0.3$ (Fig.~\ref{fig2}(h)), the striped and uneven $n_k$ distribution characterizes CDW. This confirms that the interlayer momentum shift can drive a phase transition from the fractional topological phase to a CDW.


\textit{cTBG with stretched lattice---.} The other example of FCI is chiral twisted bilayer graphene at certain magic twist angles~\cite{tarnopolsky2019origin}. The chiral limit of cTBG is realized by zeroing the intrasublattice tunneling, in which case the energy dispersion at charge neutrality becomes exactly flat at magic twist angles~\cite{tarnopolsky2019origin}. It has been established that the wavefunction of cTBG exhibits a LLL character in the chiral limit, and can be expressed in terms of LLL wavefunctions~\cite{wang2021chiral,wang2021exact},
\begin{equation}
\Phi_{\mathbf{k}}(\mathbf{r})
=
\mathcal{N}_{\mathbf{k}}
\mathcal{B}(\mathbf{r})
\psi_{\mathbf{k}}(\mathbf{r}),
\end{equation}
where $\mathcal{N}_k$  is a normalization factor, $\mathcal{B}(\bm{r})$ is a ${k}$-independent quasiperiodic function, and $\psi_{{k}}(\bm{r})$ denotes the LLL wavefunction, with the magnetic length $\ell_B=\sqrt{\hbar/eB}=1$ hereafter. In the symmetric gauge, $\psi_{{k}}(\bm{r})$ takes the form $\psi_{{k}}(\bm{r}) = \sigma(z + ik) \, \exp(ik^*z) \, \exp\!\left(-\frac{1}{2}|z|^2 - \frac{1}{2}|k|^2\right)$,
where \(\sigma(z)\) is the modified Weierstrass sigma function~\cite{ferrari1990two,ferrari1995wannier,haldane2018modular,wang2019lattice}, and $z \equiv \omega_a r^a$.  And  $|\mathcal{B}(\bm{r})|^2=w_0+w_1\sum_{\bm{b}}e^{i\bm{b}\cdot \bm{r}}$, where $\bm{b}$ includes $\pm\bm{b}_1$, $\pm\bm{b}_2$ and $\pm(\bm{b}_1+\bm{b}_2)$, in which $\bm{b}_1$ and $\bm{b}_2$ are the reciprocal lattice vectors. We start from the simplest case where the graphene is a regular hexagon, adapting the parameters in ref.~\cite{wang2021exact}. This model works for $\frac{w_1}{w_0}\leq 1/3$~\cite{supple}, and we take $w_0=0.89$, $w_1=0.216$ hereafter. 

\begin{figure}
\centering
\includegraphics[width=1.0 \linewidth]{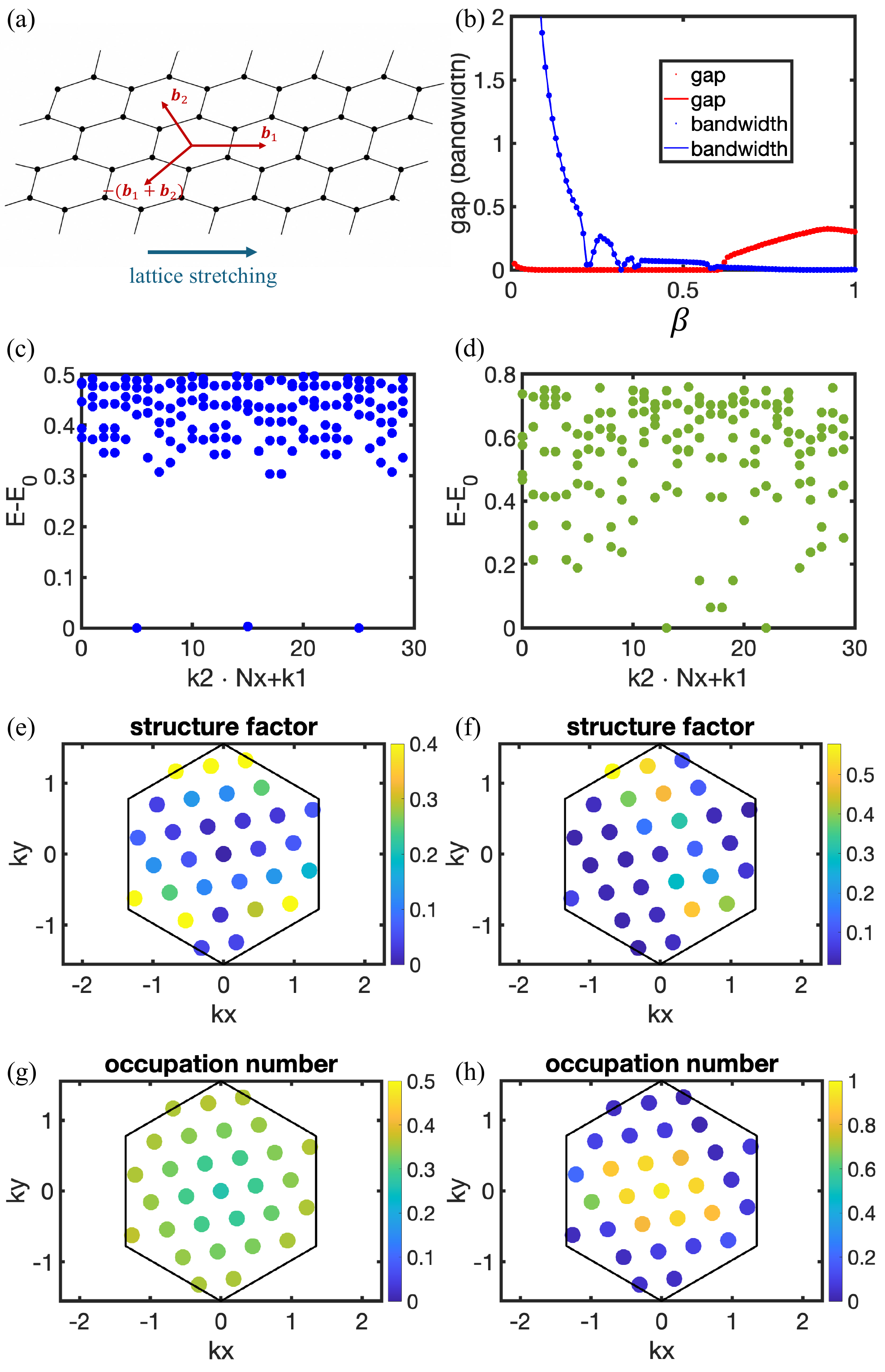}
\caption{(a) The schematic of stretching the lattice to induce anisotropy. (b) FCI gap and ground state bandwidth as functions of lattice ratio $\beta$. (c,d) The energy spectra of the Coulomb interaction on the ideal flat band corresponding to $\beta=1$ and $\beta=0.5$, respectively. (e,f) Structure factor of the ground states under the Coulomb interaction projected onto the ideal flat band corresponding to a regular honeycomb lattice ($\beta=1$) and a stretched lattice ($\beta=0.5$), respectively. (g,h) Orbital occupation number of the ground state for $\beta = 1$ and $\beta = 0.5$, respectively.
}
\label{fig3}
\end{figure}

As shown in Fig.~\ref{fig3}(a), anisotropy is induced by lattice stretching, which alters the aspect ratio of the hexagonal lattice and consequently modifies the shape of the reciprocal lattice unit cell. We define the anisotropy parameter $\beta = |\bm{b}_1| / |\bm{b}_2|$. Specifically, the regular hexagon corresponds to $\beta = 1$, and the deviation of $\beta$ from unity quantifies the degree of anisotropy. We now project the Coulomb interaction onto the ideal flat band, yielding an expression formally identical to Eq.~\eqref{twobody} with $V(\bm{q}) = 2\pi/q$ (in Gaussian units, where the dielectric constant has been absorbed).


As the lattice anisotropy increases, the system evolves from an incompressible topological phase (FCI) to a compressible Fermi-liquid state. Owing to the threefold center-of-mass degeneracy, the ground-state manifold exhibits a finite bandwidth; accordingly, the energy separation between the third and fourth lowest eigenstates is identified as the FCI gap. The evolution of both the FCI gap and the ground-state bandwidth is shown in Fig.~\ref{fig3}(b). In the topological phase regime ($\beta$ close to unity), the ground-state manifold remains nearly degenerate, exhibiting a small bandwidth that is well separated from higher-energy states by a finite incompressibility gap (see Fig.~\ref{fig3}(b)). Consistently, the structure factor is suppressed at small momentum and has a ring peak (see Fig.~\ref{fig3}(e)). As shown in Figs.~\ref{fig3}(g), the occupation number is uniformly distributed, lacking any well-defined Fermi surface. As the anisotropy is further increased (i.e., with decreasing $\beta$), the ground-state manifold bandwidth progressively broadens and eventually exceeds the FCI gap, indicating destabilization of the FCI phase (see Fig.~\ref{fig3}(b)). In the strongly anisotropic regime ($\beta \sim 0.5$ or smaller), the degeneracy of the three-manifold is lifted and the gap closes, indicating the vanishing of the incompressible FCI phase. In addition, as shown in Fig.~\ref{fig3}(h), the occupation number develops a sharp momentum-space boundary consistent with the emergence of a Fermi-surface-like contour within the Brillouin zone. Together, these signatures provide clear evidence for an anisotropy-driven transition, induced by lattice stretching, from an incompressible topological phase to a compressible Fermi liquid. Interestingly, the mapping between the ideal flat band and the LLL guarantees the existence of exact zero-energy ground states for short-range repulsive generalized pseudopotentials. This suggests the robustness of $V_1$ against lattice scratching anisotropy and that the corresponding phase transitions for the $V_1$ do not occur at the same anisotropy scale as those driven by the Coulomb interaction~\cite{wang2021exact}.

\textit{LL correspondence with anisotropic effective mass.---}FCI shares many essential features with fractional quantum Hall states, yet the lattice structure and nonuniform quantum geometry of Chern bands can give rise to phenomena beyond those of continuum LLs. Thus, it is interesting to investigate the anisotropy on a single LL as a comparison of the anisotropic topological state and its phase transition, helping uncover the mechanisms underlying the stabilized topological phases.

We start from the single-particle kinetic Hamiltonian with an anisotropic effective mass~\cite{yang2012band},
\begin{equation}
H=\frac{1}{2m}g_m^{ab}(p_a-eA_a)(p_b-eA_b),
\end{equation}
where $g_m^{ab}$ is the mass metric. With the magnetic length $\ell_B=\sqrt{\hbar/{eB}}$, two sets of spatial coordinates, cyclotron and guiding center coordinates are defined as: $\eta^a=-\epsilon^{ab}\ell_B^2(p_b-eA_b), R^a=r^a-\eta^a,$ and the cyclotron and guiding-center ladder operators are: $a^\dagger= \frac{1}{\sqrt{2}\ell_B} \left( \sqrt{g_m^{xx}}\,\eta^x -i\sqrt{g_m^{yy}}\,\eta^y \right),$ and $b^\dagger= \frac{1}{\sqrt{2}\ell_B} \left( \sqrt{g_g^{xx}}\,R^x +i\sqrt{g_g^{yy}}\,R^y \right),$ where $g_m$ and $g_g$ encode the mass and guiding-center metrics, respectively. The two sets of ladder operators remain decoupled, $[a,a^\dagger]=[b,b^\dagger]=1, [a,b]=[a,b^\dagger]=0$. 

The rotational invariance only exists if the cyclotron metric and the guiding center metric are congruent, that is, $g_m=g_g$, which is a special case generally adopted in the literature for technical convenience. 
Moreover, one can also consider an anisotropic Hamiltonian $V(\bm{q})$, which normally is a function of $|\bm{q}|$ and $|\bm{q}|=\sqrt{g^{ab}_iq_aq_b}$. Here $a, b = x,y$ denote Cartesian coordinates, and the Einstein summation convention is implied. When the Hamiltonian anisotropy tensor is taken as the same as the effective mass tensor, that is $g_i=g_m$, families of anisotropic fractional quantum Hall states can be constructed, as the exact zero energy ground states of appropriate anisotropic short-range two- or multi-particle interactions~\cite{Qiu2012modelanisotropicH,yang2017generalized}. These states therefore, provide an explicit manifestation of the geometric degree of freedom in the fractional quantum Hall effect, and indicate the robustness against the anisotropy of the system.

Moreover, $g_i=g_m$ is not required to protect the topological phases of the FQHE~\cite{Haldane2011anisotropygeometry}.  Physically, only the relative difference between these metrics matters. Thus, without loss of generality, we take $g_i=\delta^{ab}$ and $g_g=\delta^{ab}$. The simplest case of the effective mass anisotropy reads as,
\begin{equation}
    g_m=\left(
    \begin{matrix}
        1/\alpha & 0 \\
        0 & \alpha
    \end{matrix}
    \right),
\end{equation}
in which $\alpha$ is the anisotropy parameter, and $\alpha \in (0,1]$ characterizes the strength of the anisotropy. Specifically, $\alpha=1$ corresponds to the isotropic case, and the smaller $\alpha$ is, the larger the anisotropy. In the LLL, even though the finite-size spectra of the $V_1$ at $\nu=1/3$ exhibit an apparent gap-closing point as the anisotropy increases, finite-size scaling shows that this critical point extrapolates to $\alpha = 0$ in the thermodynamic limit. This implies that the gap closing is merely a finite-size effect and demonstrates the robustness of the anisotropic Laughlin phase~\cite{supple}.

\begin{figure}
\centering
\includegraphics[width=1.0\linewidth]{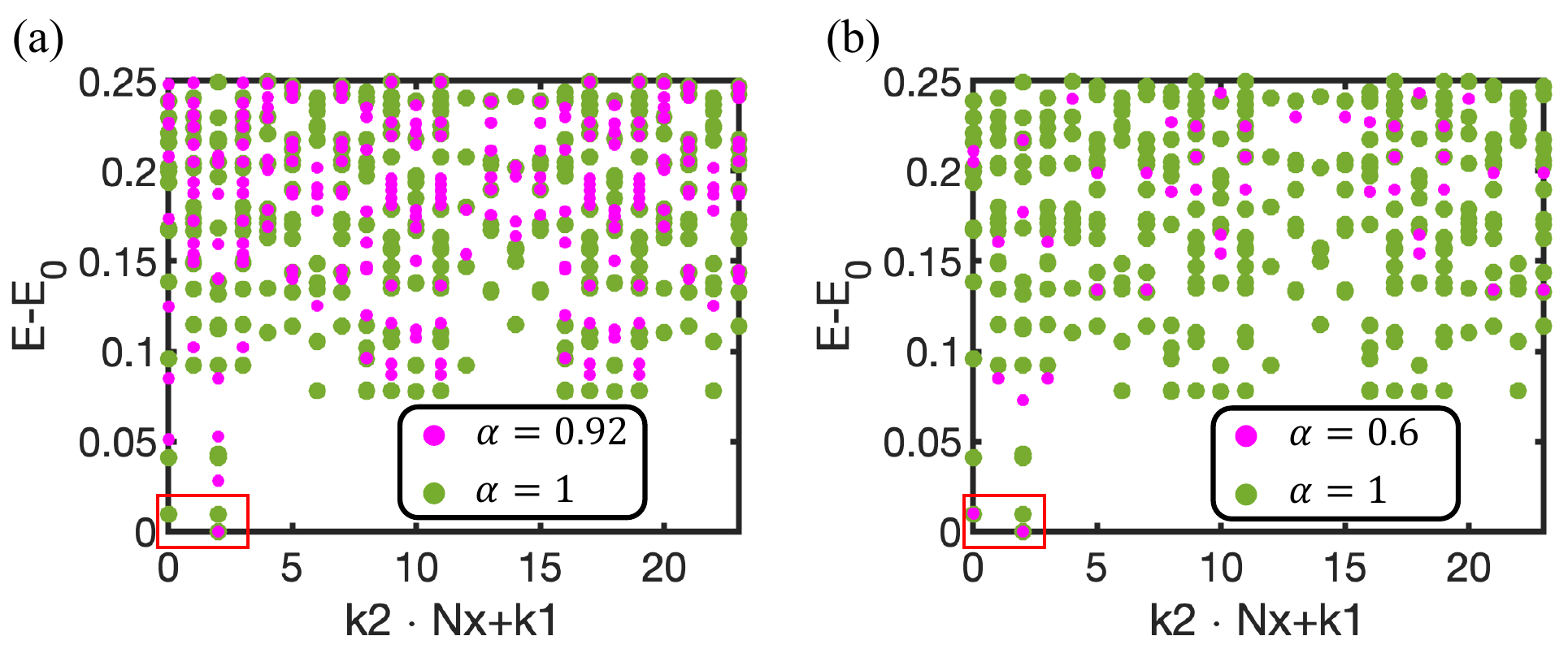}
\caption{ The spectra of Coulomb interaction on the first Landau Level at filling $\nu=1/2$ for (a) $g_i=\delta^{ab}$ with $\alpha=1$ (green dots) and $\alpha=0.92$ (magenta dots). (b) $g_i=g_m$ with $\alpha=1$ (green dots) and $\alpha=0.6$ (magenta dots). The $\alpha=1$ corresponds to the isotropic case, giving the sixfold Moore-Read states in the red boxes.
}
\label{fig4}
\end{figure}

However, higher LLs are particularly susceptible to interaction-driven competing phases under strong anisotropy, including charge-density-wave, stripe/nematic, and bubble orders~\cite{yang2012band,you2014theory,shibata2001ground,yang2023cascade}. 
For example, it has been systematically shown that in the first LL, the Coulomb interaction can stabilize the Moore-Read state at filling factor $\nu=1/2$\cite{moore1991nonabelions,Rezayi2000Coulomb}. This is a non-Abelian topological phase characterized by a threefold topological ground-state degeneracy on the torus, in addition to the universal twofold center-of-mass degeneracy~\cite{Bergholtz2006threefold}. As shown in Fig.~\ref{fig4}(a), for a system with $N_e = 12$, $N_x = 4$, $N_y = 6$, the sixfold ground-state manifold is distributed between the momentum sectors $(0,0)$ and $(2,0)$, with two and four states residing in each sector, respectively.

While the system appears more stable in the vicinity of the isotropic point $\alpha=1$, the Moore-Read phase persists with slight effective mass anisotropy. Nevertheless, the topological phase becomes unstable as $\alpha$ deviates far from unity, as evidenced by the redistribution of the ground states away from their original momentum sectors, as shown in Fig.~\ref{fig4}(a) in which $\alpha=0.92$. One can observe that the markedly reduced energy separation between the third and fourth lowest-energy states (despite center-of-mass degeneracy) indicates the collapse of the FQH gap. In addition, the anisotropic Moore-Read state characterized by $g_i=g_m=g_g\neq \delta^{ab}$ exhibits greater robustness than the case in which only $g_m\neq \delta^{ab}$. In this case, a phase transition from FQH to charge-ordered phase occurs around a critical point $\alpha\approx0.7$, below which the Moore-Read gap is smaller than the ground state bandwidth, and the comparison of the isotropic Coulomb interaction and $\alpha=0.6$ case is shown in Fig.~\ref{fig4}(b).



\emph{Conclusion and discussion.---}In this work, we demonstrated the robustness of FCI phase against temperate anisotrpy and that strong anisotropy in FCIs provides a generic mechanism for destabilizing incompressible topological liquids such as moir\'{e} flat bands. By considering heterostrain induced by interlayer momentum shift, and lattice deformation in ideal flat bands relevant to cTBG, we uncovered a unified physical picture in which anisotropy drives correlated topological phases toward qualitatively distinct competing orders, including CDW order and compressible Fermi-liquid-like states. More specifically, strong anisotropy reshapes the band structure and projected interaction of the partially filled Chern band, destabilizing the incompressible FCI state in favor of either stripe-type CDW order through translational-symmetry breaking or a compressible Fermi-liquid-like phase with Fermi-surface-like momentum-space occupations.

The anisotropy-driven phase transitions we have established in FCI models find strong parallels in recent experimental observations across multiple moir\'{e} platforms. Our theoretical and numerical analyses identify heterostrain as a key parameter that induces interlayer momentum shifts and destabilizes FCIs, particularly at small twist angles. We estimate a critical heterostrain of $\epsilon_c \approx 4\%$ at which the FCI-CDW transition occurs in twisted MoTe$_2$ at $\theta = 3.8^\circ$. This means the FCI phase is robust against anisotropy, as typical strain values in experiments are less than $2\%$~\cite{hu2022nonlinear,huang2023giant,zhang2022giant}. Nevertheless, we anticipate that in other moir\'{e} TMDs such as twisted MoS$_2$~\cite{zhou2022moire}, this phase transition may be observable in future experiments. 


\emph{Acknowledgement.---}We would like to acknowledge insightful discussions with Bo Yang and Kun Yang. This work is supported by the Singapore Ministry of Education (MOE) Academic Research Fund Tier 3 Grant (No. MOE-MOET32023-0003) “Quantum Geometric Advantage”, and Singapore Ministry of Education (MOE) Academic Research Fund Tier 2 Grant (No. MOE-T2EP50124-0017).


\begin{thebibliography}{58}%
\makeatletter
\providecommand \@ifxundefined [1]{%
 \@ifx{#1\undefined}
}%
\providecommand \@ifnum [1]{%
 \ifnum #1\expandafter \@firstoftwo
 \else \expandafter \@secondoftwo
 \fi
}%
\providecommand \@ifx [1]{%
 \ifx #1\expandafter \@firstoftwo
 \else \expandafter \@secondoftwo
 \fi
}%
\providecommand \natexlab [1]{#1}%
\providecommand \enquote  [1]{``#1''}%
\providecommand \bibnamefont  [1]{#1}%
\providecommand \bibfnamefont [1]{#1}%
\providecommand \citenamefont [1]{#1}%
\providecommand \href@noop [0]{\@secondoftwo}%
\providecommand \href [0]{\begingroup \@sanitize@url \@href}%
\providecommand \@href[1]{\@@startlink{#1}\@@href}%
\providecommand \@@href[1]{\endgroup#1\@@endlink}%
\providecommand \@sanitize@url [0]{\catcode `\\12\catcode `\$12\catcode
  `\&12\catcode `\#12\catcode `\^12\catcode `\_12\catcode `\%12\relax}%
\providecommand \@@startlink[1]{}%
\providecommand \@@endlink[0]{}%
\providecommand \url  [0]{\begingroup\@sanitize@url \@url }%
\providecommand \@url [1]{\endgroup\@href {#1}{\urlprefix }}%
\providecommand \urlprefix  [0]{URL }%
\providecommand \Eprint [0]{\href }%
\providecommand \doibase [0]{http://dx.doi.org/}%
\providecommand \selectlanguage [0]{\@gobble}%
\providecommand \bibinfo  [0]{\@secondoftwo}%
\providecommand \bibfield  [0]{\@secondoftwo}%
\providecommand \translation [1]{[#1]}%
\providecommand \BibitemOpen [0]{}%
\providecommand \bibitemStop [0]{}%
\providecommand \bibitemNoStop [0]{.\EOS\space}%
\providecommand \EOS [0]{\spacefactor3000\relax}%
\providecommand \BibitemShut  [1]{\csname bibitem#1\endcsname}%
\let\auto@bib@innerbib\@empty
\bibitem [{\citenamefont {Fradkin}(2013)}]{fradkin2013field}%
  \BibitemOpen
  \bibfield  {author} {\bibinfo {author} {\bibfnamefont {E.}~\bibnamefont
  {Fradkin}},\ }\href@noop {} {\emph {\bibinfo {title} {Field theories of
  condensed matter physics}}}\ (\bibinfo  {publisher} {Cambridge University
  Press},\ \bibinfo {year} {2013})\BibitemShut {NoStop}%
\bibitem [{\citenamefont {Stormer}\ \emph {et~al.}(1999)\citenamefont
  {Stormer}, \citenamefont {Tsui},\ and\ \citenamefont
  {Gossard}}]{stormer1999fractional}%
  \BibitemOpen
  \bibfield  {author} {\bibinfo {author} {\bibfnamefont {H.~L.}\ \bibnamefont
  {Stormer}}, \bibinfo {author} {\bibfnamefont {D.~C.}\ \bibnamefont {Tsui}}, \
  and\ \bibinfo {author} {\bibfnamefont {A.~C.}\ \bibnamefont {Gossard}},\
  }\href@noop {} {\bibfield  {journal} {\bibinfo  {journal} {Reviews of Modern
  Physics}\ }\textbf {\bibinfo {volume} {71}},\ \bibinfo {pages} {S298}
  (\bibinfo {year} {1999})}\BibitemShut {NoStop}%
\bibitem [{\citenamefont {Stormer}(1999)}]{stormer1999nobel}%
  \BibitemOpen
  \bibfield  {author} {\bibinfo {author} {\bibfnamefont {H.~L.}\ \bibnamefont
  {Stormer}},\ }\href@noop {} {\bibfield  {journal} {\bibinfo  {journal}
  {Reviews of Modern Physics}\ }\textbf {\bibinfo {volume} {71}},\ \bibinfo
  {pages} {875} (\bibinfo {year} {1999})}\BibitemShut {NoStop}%
\bibitem [{\citenamefont {Moore}\ and\ \citenamefont
  {Read}(1991)}]{moore1991nonabelions}%
  \BibitemOpen
  \bibfield  {author} {\bibinfo {author} {\bibfnamefont {G.}~\bibnamefont
  {Moore}}\ and\ \bibinfo {author} {\bibfnamefont {N.}~\bibnamefont {Read}},\
  }\href@noop {} {\bibfield  {journal} {\bibinfo  {journal} {Nuclear Physics
  B}\ }\textbf {\bibinfo {volume} {360}},\ \bibinfo {pages} {362} (\bibinfo
  {year} {1991})}\BibitemShut {NoStop}%
\bibitem [{\citenamefont {Bolotin}\ \emph {et~al.}(2009)\citenamefont
  {Bolotin}, \citenamefont {Ghahari}, \citenamefont {Shulman}, \citenamefont
  {Stormer},\ and\ \citenamefont {Kim}}]{bolotin2009observation}%
  \BibitemOpen
  \bibfield  {author} {\bibinfo {author} {\bibfnamefont {K.~I.}\ \bibnamefont
  {Bolotin}}, \bibinfo {author} {\bibfnamefont {F.}~\bibnamefont {Ghahari}},
  \bibinfo {author} {\bibfnamefont {M.~D.}\ \bibnamefont {Shulman}}, \bibinfo
  {author} {\bibfnamefont {H.~L.}\ \bibnamefont {Stormer}}, \ and\ \bibinfo
  {author} {\bibfnamefont {P.}~\bibnamefont {Kim}},\ }\href@noop {} {\bibfield
  {journal} {\bibinfo  {journal} {Nature}\ }\textbf {\bibinfo {volume} {462}},\
  \bibinfo {pages} {196} (\bibinfo {year} {2009})}\BibitemShut {NoStop}%
\bibitem [{\citenamefont {Cage}\ \emph {et~al.}(2012)\citenamefont {Cage},
  \citenamefont {Klitzing}, \citenamefont {Chang}, \citenamefont {Duncan},
  \citenamefont {Haldane}, \citenamefont {Laughlin}, \citenamefont {Pruisken},\
  and\ \citenamefont {Thouless}}]{cage2012quantum}%
  \BibitemOpen
  \bibfield  {author} {\bibinfo {author} {\bibfnamefont {M.~E.}\ \bibnamefont
  {Cage}}, \bibinfo {author} {\bibfnamefont {K.}~\bibnamefont {Klitzing}},
  \bibinfo {author} {\bibfnamefont {A.}~\bibnamefont {Chang}}, \bibinfo
  {author} {\bibfnamefont {F.}~\bibnamefont {Duncan}}, \bibinfo {author}
  {\bibfnamefont {M.}~\bibnamefont {Haldane}}, \bibinfo {author} {\bibfnamefont
  {R.~B.}\ \bibnamefont {Laughlin}}, \bibinfo {author} {\bibfnamefont
  {A.}~\bibnamefont {Pruisken}}, \ and\ \bibinfo {author} {\bibfnamefont
  {D.}~\bibnamefont {Thouless}},\ }\href@noop {} {\emph {\bibinfo {title} {The
  quantum Hall effect}}}\ (\bibinfo  {publisher} {Springer Science \& Business
  Media},\ \bibinfo {year} {2012})\BibitemShut {NoStop}%
\bibitem [{\citenamefont {Ju}\ \emph {et~al.}(2024)\citenamefont {Ju},
  \citenamefont {MacDonald}, \citenamefont {Mak}, \citenamefont {Shan},\ and\
  \citenamefont {Xu}}]{ju2024fractional}%
  \BibitemOpen
  \bibfield  {author} {\bibinfo {author} {\bibfnamefont {L.}~\bibnamefont
  {Ju}}, \bibinfo {author} {\bibfnamefont {A.~H.}\ \bibnamefont {MacDonald}},
  \bibinfo {author} {\bibfnamefont {K.~F.}\ \bibnamefont {Mak}}, \bibinfo
  {author} {\bibfnamefont {J.}~\bibnamefont {Shan}}, \ and\ \bibinfo {author}
  {\bibfnamefont {X.}~\bibnamefont {Xu}},\ }\href@noop {} {\bibfield  {journal}
  {\bibinfo  {journal} {Nature Reviews Materials}\ }\textbf {\bibinfo {volume}
  {9}},\ \bibinfo {pages} {455} (\bibinfo {year} {2024})}\BibitemShut {NoStop}%
\bibitem [{\citenamefont {Bistritzer}\ and\ \citenamefont
  {MacDonald}(2011)}]{bistritzer2011moire}%
  \BibitemOpen
  \bibfield  {author} {\bibinfo {author} {\bibfnamefont {R.}~\bibnamefont
  {Bistritzer}}\ and\ \bibinfo {author} {\bibfnamefont {A.~H.}\ \bibnamefont
  {MacDonald}},\ }\href@noop {} {\bibfield  {journal} {\bibinfo  {journal}
  {Proceedings of the National Academy of Sciences}\ }\textbf {\bibinfo
  {volume} {108}},\ \bibinfo {pages} {12233} (\bibinfo {year}
  {2011})}\BibitemShut {NoStop}%
\bibitem [{\citenamefont {Cai}\ \emph {et~al.}(2023)\citenamefont {Cai},
  \citenamefont {Anderson}, \citenamefont {Wang}, \citenamefont {Zhang},
  \citenamefont {Liu}, \citenamefont {Holtzmann}, \citenamefont {Zhang},
  \citenamefont {Fan}, \citenamefont {Taniguchi}, \citenamefont {Watanabe}
  \emph {et~al.}}]{cai2023signatures}%
  \BibitemOpen
  \bibfield  {author} {\bibinfo {author} {\bibfnamefont {J.}~\bibnamefont
  {Cai}}, \bibinfo {author} {\bibfnamefont {E.}~\bibnamefont {Anderson}},
  \bibinfo {author} {\bibfnamefont {C.}~\bibnamefont {Wang}}, \bibinfo {author}
  {\bibfnamefont {X.}~\bibnamefont {Zhang}}, \bibinfo {author} {\bibfnamefont
  {X.}~\bibnamefont {Liu}}, \bibinfo {author} {\bibfnamefont {W.}~\bibnamefont
  {Holtzmann}}, \bibinfo {author} {\bibfnamefont {Y.}~\bibnamefont {Zhang}},
  \bibinfo {author} {\bibfnamefont {F.}~\bibnamefont {Fan}}, \bibinfo {author}
  {\bibfnamefont {T.}~\bibnamefont {Taniguchi}}, \bibinfo {author}
  {\bibfnamefont {K.}~\bibnamefont {Watanabe}},  \emph {et~al.},\ }\href@noop
  {} {\bibfield  {journal} {\bibinfo  {journal} {Nature}\ }\textbf {\bibinfo
  {volume} {622}},\ \bibinfo {pages} {63} (\bibinfo {year} {2023})}\BibitemShut
  {NoStop}%
\bibitem [{\citenamefont {Park}\ \emph {et~al.}(2023)\citenamefont {Park},
  \citenamefont {Cai}, \citenamefont {Anderson}, \citenamefont {Zhang},
  \citenamefont {Zhu}, \citenamefont {Liu}, \citenamefont {Wang}, \citenamefont
  {Holtzmann}, \citenamefont {Hu}, \citenamefont {Liu} \emph
  {et~al.}}]{park2023observation}%
  \BibitemOpen
  \bibfield  {author} {\bibinfo {author} {\bibfnamefont {H.}~\bibnamefont
  {Park}}, \bibinfo {author} {\bibfnamefont {J.}~\bibnamefont {Cai}}, \bibinfo
  {author} {\bibfnamefont {E.}~\bibnamefont {Anderson}}, \bibinfo {author}
  {\bibfnamefont {Y.}~\bibnamefont {Zhang}}, \bibinfo {author} {\bibfnamefont
  {J.}~\bibnamefont {Zhu}}, \bibinfo {author} {\bibfnamefont {X.}~\bibnamefont
  {Liu}}, \bibinfo {author} {\bibfnamefont {C.}~\bibnamefont {Wang}}, \bibinfo
  {author} {\bibfnamefont {W.}~\bibnamefont {Holtzmann}}, \bibinfo {author}
  {\bibfnamefont {C.}~\bibnamefont {Hu}}, \bibinfo {author} {\bibfnamefont
  {Z.}~\bibnamefont {Liu}},  \emph {et~al.},\ }\href@noop {} {\bibfield
  {journal} {\bibinfo  {journal} {Nature}\ }\textbf {\bibinfo {volume} {622}},\
  \bibinfo {pages} {74} (\bibinfo {year} {2023})}\BibitemShut {NoStop}%
\bibitem [{\citenamefont {Zeng}\ \emph {et~al.}(2023)\citenamefont {Zeng},
  \citenamefont {Xia}, \citenamefont {Kang}, \citenamefont {Zhu}, \citenamefont
  {Kn{\"u}ppel}, \citenamefont {Vaswani}, \citenamefont {Watanabe},
  \citenamefont {Taniguchi}, \citenamefont {Mak},\ and\ \citenamefont
  {Shan}}]{zeng2023thermodynamic}%
  \BibitemOpen
  \bibfield  {author} {\bibinfo {author} {\bibfnamefont {Y.}~\bibnamefont
  {Zeng}}, \bibinfo {author} {\bibfnamefont {Z.}~\bibnamefont {Xia}}, \bibinfo
  {author} {\bibfnamefont {K.}~\bibnamefont {Kang}}, \bibinfo {author}
  {\bibfnamefont {J.}~\bibnamefont {Zhu}}, \bibinfo {author} {\bibfnamefont
  {P.}~\bibnamefont {Kn{\"u}ppel}}, \bibinfo {author} {\bibfnamefont
  {C.}~\bibnamefont {Vaswani}}, \bibinfo {author} {\bibfnamefont
  {K.}~\bibnamefont {Watanabe}}, \bibinfo {author} {\bibfnamefont
  {T.}~\bibnamefont {Taniguchi}}, \bibinfo {author} {\bibfnamefont {K.~F.}\
  \bibnamefont {Mak}}, \ and\ \bibinfo {author} {\bibfnamefont
  {J.}~\bibnamefont {Shan}},\ }\href@noop {} {\bibfield  {journal} {\bibinfo
  {journal} {Nature}\ }\textbf {\bibinfo {volume} {622}},\ \bibinfo {pages}
  {69} (\bibinfo {year} {2023})}\BibitemShut {NoStop}%
\bibitem [{\citenamefont {Xu}\ \emph {et~al.}(2023)\citenamefont {Xu},
  \citenamefont {Sun}, \citenamefont {Jia}, \citenamefont {Liu}, \citenamefont
  {Xu}, \citenamefont {Li}, \citenamefont {Gu}, \citenamefont {Watanabe},
  \citenamefont {Taniguchi}, \citenamefont {Tong} \emph
  {et~al.}}]{xu2023observation}%
  \BibitemOpen
  \bibfield  {author} {\bibinfo {author} {\bibfnamefont {F.}~\bibnamefont
  {Xu}}, \bibinfo {author} {\bibfnamefont {Z.}~\bibnamefont {Sun}}, \bibinfo
  {author} {\bibfnamefont {T.}~\bibnamefont {Jia}}, \bibinfo {author}
  {\bibfnamefont {C.}~\bibnamefont {Liu}}, \bibinfo {author} {\bibfnamefont
  {C.}~\bibnamefont {Xu}}, \bibinfo {author} {\bibfnamefont {C.}~\bibnamefont
  {Li}}, \bibinfo {author} {\bibfnamefont {Y.}~\bibnamefont {Gu}}, \bibinfo
  {author} {\bibfnamefont {K.}~\bibnamefont {Watanabe}}, \bibinfo {author}
  {\bibfnamefont {T.}~\bibnamefont {Taniguchi}}, \bibinfo {author}
  {\bibfnamefont {B.}~\bibnamefont {Tong}},  \emph {et~al.},\ }\href@noop {}
  {\bibfield  {journal} {\bibinfo  {journal} {Physical Review X}\ }\textbf
  {\bibinfo {volume} {13}},\ \bibinfo {pages} {031037} (\bibinfo {year}
  {2023})}\BibitemShut {NoStop}%
\bibitem [{\citenamefont {Redekop}\ \emph {et~al.}(2024)\citenamefont
  {Redekop}, \citenamefont {Zhang}, \citenamefont {Park}, \citenamefont {Cai},
  \citenamefont {Anderson}, \citenamefont {Sheekey}, \citenamefont {Arp},
  \citenamefont {Babikyan}, \citenamefont {Salters}, \citenamefont {Watanabe}
  \emph {et~al.}}]{redekop2024direct}%
  \BibitemOpen
  \bibfield  {author} {\bibinfo {author} {\bibfnamefont {E.}~\bibnamefont
  {Redekop}}, \bibinfo {author} {\bibfnamefont {C.}~\bibnamefont {Zhang}},
  \bibinfo {author} {\bibfnamefont {H.}~\bibnamefont {Park}}, \bibinfo {author}
  {\bibfnamefont {J.}~\bibnamefont {Cai}}, \bibinfo {author} {\bibfnamefont
  {E.}~\bibnamefont {Anderson}}, \bibinfo {author} {\bibfnamefont
  {O.}~\bibnamefont {Sheekey}}, \bibinfo {author} {\bibfnamefont
  {T.}~\bibnamefont {Arp}}, \bibinfo {author} {\bibfnamefont {G.}~\bibnamefont
  {Babikyan}}, \bibinfo {author} {\bibfnamefont {S.}~\bibnamefont {Salters}},
  \bibinfo {author} {\bibfnamefont {K.}~\bibnamefont {Watanabe}},  \emph
  {et~al.},\ }\href@noop {} {\bibfield  {journal} {\bibinfo  {journal}
  {Nature}\ }\textbf {\bibinfo {volume} {635}},\ \bibinfo {pages} {584}
  (\bibinfo {year} {2024})}\BibitemShut {NoStop}%
\bibitem [{\citenamefont {Lu}\ \emph {et~al.}(2024)\citenamefont {Lu},
  \citenamefont {Han}, \citenamefont {Yao}, \citenamefont {Reddy},
  \citenamefont {Yang}, \citenamefont {Seo}, \citenamefont {Watanabe},
  \citenamefont {Taniguchi}, \citenamefont {Fu},\ and\ \citenamefont
  {Ju}}]{lu2024fractional}%
  \BibitemOpen
  \bibfield  {author} {\bibinfo {author} {\bibfnamefont {Z.}~\bibnamefont
  {Lu}}, \bibinfo {author} {\bibfnamefont {T.}~\bibnamefont {Han}}, \bibinfo
  {author} {\bibfnamefont {Y.}~\bibnamefont {Yao}}, \bibinfo {author}
  {\bibfnamefont {A.~P.}\ \bibnamefont {Reddy}}, \bibinfo {author}
  {\bibfnamefont {J.}~\bibnamefont {Yang}}, \bibinfo {author} {\bibfnamefont
  {J.}~\bibnamefont {Seo}}, \bibinfo {author} {\bibfnamefont {K.}~\bibnamefont
  {Watanabe}}, \bibinfo {author} {\bibfnamefont {T.}~\bibnamefont {Taniguchi}},
  \bibinfo {author} {\bibfnamefont {L.}~\bibnamefont {Fu}}, \ and\ \bibinfo
  {author} {\bibfnamefont {L.}~\bibnamefont {Ju}},\ }\href@noop {} {\bibfield
  {journal} {\bibinfo  {journal} {Nature}\ }\textbf {\bibinfo {volume} {626}},\
  \bibinfo {pages} {759} (\bibinfo {year} {2024})}\BibitemShut {NoStop}%
\bibitem [{\citenamefont {Xie}\ \emph {et~al.}(2021)\citenamefont {Xie},
  \citenamefont {Pierce}, \citenamefont {Park}, \citenamefont {Parker},
  \citenamefont {Khalaf}, \citenamefont {Ledwith}, \citenamefont {Cao},
  \citenamefont {Lee}, \citenamefont {Chen}, \citenamefont {Forrester} \emph
  {et~al.}}]{xie2021fractional}%
  \BibitemOpen
  \bibfield  {author} {\bibinfo {author} {\bibfnamefont {Y.}~\bibnamefont
  {Xie}}, \bibinfo {author} {\bibfnamefont {A.~T.}\ \bibnamefont {Pierce}},
  \bibinfo {author} {\bibfnamefont {J.~M.}\ \bibnamefont {Park}}, \bibinfo
  {author} {\bibfnamefont {D.~E.}\ \bibnamefont {Parker}}, \bibinfo {author}
  {\bibfnamefont {E.}~\bibnamefont {Khalaf}}, \bibinfo {author} {\bibfnamefont
  {P.}~\bibnamefont {Ledwith}}, \bibinfo {author} {\bibfnamefont
  {Y.}~\bibnamefont {Cao}}, \bibinfo {author} {\bibfnamefont {S.~H.}\
  \bibnamefont {Lee}}, \bibinfo {author} {\bibfnamefont {S.}~\bibnamefont
  {Chen}}, \bibinfo {author} {\bibfnamefont {P.~R.}\ \bibnamefont {Forrester}},
   \emph {et~al.},\ }\href@noop {} {\bibfield  {journal} {\bibinfo  {journal}
  {Nature}\ }\textbf {\bibinfo {volume} {600}},\ \bibinfo {pages} {439}
  (\bibinfo {year} {2021})}\BibitemShut {NoStop}%
\bibitem [{\citenamefont {Xie}\ \emph {et~al.}(2025)\citenamefont {Xie},
  \citenamefont {Huo}, \citenamefont {Lu}, \citenamefont {Feng}, \citenamefont
  {Zhang}, \citenamefont {Wang}, \citenamefont {Yang}, \citenamefont
  {Watanabe}, \citenamefont {Taniguchi}, \citenamefont {Liu} \emph
  {et~al.}}]{xie2025tunable}%
  \BibitemOpen
  \bibfield  {author} {\bibinfo {author} {\bibfnamefont {J.}~\bibnamefont
  {Xie}}, \bibinfo {author} {\bibfnamefont {Z.}~\bibnamefont {Huo}}, \bibinfo
  {author} {\bibfnamefont {X.}~\bibnamefont {Lu}}, \bibinfo {author}
  {\bibfnamefont {Z.}~\bibnamefont {Feng}}, \bibinfo {author} {\bibfnamefont
  {Z.}~\bibnamefont {Zhang}}, \bibinfo {author} {\bibfnamefont
  {W.}~\bibnamefont {Wang}}, \bibinfo {author} {\bibfnamefont {Q.}~\bibnamefont
  {Yang}}, \bibinfo {author} {\bibfnamefont {K.}~\bibnamefont {Watanabe}},
  \bibinfo {author} {\bibfnamefont {T.}~\bibnamefont {Taniguchi}}, \bibinfo
  {author} {\bibfnamefont {K.}~\bibnamefont {Liu}},  \emph {et~al.},\
  }\href@noop {} {\bibfield  {journal} {\bibinfo  {journal} {Nature Materials}\
  }\textbf {\bibinfo {volume} {24}},\ \bibinfo {pages} {1042} (\bibinfo {year}
  {2025})}\BibitemShut {NoStop}%
\bibitem [{\citenamefont {Aronson}\ \emph {et~al.}(2025)\citenamefont
  {Aronson}, \citenamefont {Han}, \citenamefont {Lu}, \citenamefont {Yao},
  \citenamefont {Butler}, \citenamefont {Watanabe}, \citenamefont {Taniguchi},
  \citenamefont {Ju},\ and\ \citenamefont {Ashoori}}]{aronson2025displacement}%
  \BibitemOpen
  \bibfield  {author} {\bibinfo {author} {\bibfnamefont {S.~H.}\ \bibnamefont
  {Aronson}}, \bibinfo {author} {\bibfnamefont {T.}~\bibnamefont {Han}},
  \bibinfo {author} {\bibfnamefont {Z.}~\bibnamefont {Lu}}, \bibinfo {author}
  {\bibfnamefont {Y.}~\bibnamefont {Yao}}, \bibinfo {author} {\bibfnamefont
  {J.~P.}\ \bibnamefont {Butler}}, \bibinfo {author} {\bibfnamefont
  {K.}~\bibnamefont {Watanabe}}, \bibinfo {author} {\bibfnamefont
  {T.}~\bibnamefont {Taniguchi}}, \bibinfo {author} {\bibfnamefont
  {L.}~\bibnamefont {Ju}}, \ and\ \bibinfo {author} {\bibfnamefont {R.~C.}\
  \bibnamefont {Ashoori}},\ }\href@noop {} {\bibfield  {journal} {\bibinfo
  {journal} {Physical Review X}\ }\textbf {\bibinfo {volume} {15}},\ \bibinfo
  {pages} {031026} (\bibinfo {year} {2025})}\BibitemShut {NoStop}%
\bibitem [{\citenamefont {Waters}\ \emph {et~al.}(2025)\citenamefont {Waters},
  \citenamefont {Okounkova}, \citenamefont {Su}, \citenamefont {Zhou},
  \citenamefont {Yao}, \citenamefont {Watanabe}, \citenamefont {Taniguchi},
  \citenamefont {Xu}, \citenamefont {Zhang}, \citenamefont {Folk} \emph
  {et~al.}}]{waters2025chern}%
  \BibitemOpen
  \bibfield  {author} {\bibinfo {author} {\bibfnamefont {D.}~\bibnamefont
  {Waters}}, \bibinfo {author} {\bibfnamefont {A.}~\bibnamefont {Okounkova}},
  \bibinfo {author} {\bibfnamefont {R.}~\bibnamefont {Su}}, \bibinfo {author}
  {\bibfnamefont {B.}~\bibnamefont {Zhou}}, \bibinfo {author} {\bibfnamefont
  {J.}~\bibnamefont {Yao}}, \bibinfo {author} {\bibfnamefont {K.}~\bibnamefont
  {Watanabe}}, \bibinfo {author} {\bibfnamefont {T.}~\bibnamefont {Taniguchi}},
  \bibinfo {author} {\bibfnamefont {X.}~\bibnamefont {Xu}}, \bibinfo {author}
  {\bibfnamefont {Y.-H.}\ \bibnamefont {Zhang}}, \bibinfo {author}
  {\bibfnamefont {J.}~\bibnamefont {Folk}},  \emph {et~al.},\ }\href@noop {}
  {\bibfield  {journal} {\bibinfo  {journal} {Physical Review X}\ }\textbf
  {\bibinfo {volume} {15}},\ \bibinfo {pages} {011045} (\bibinfo {year}
  {2025})}\BibitemShut {NoStop}%
\bibitem [{\citenamefont {Regnault}\ and\ \citenamefont
  {Bernevig}(2011)}]{regnault2011fractional}%
  \BibitemOpen
  \bibfield  {author} {\bibinfo {author} {\bibfnamefont {N.}~\bibnamefont
  {Regnault}}\ and\ \bibinfo {author} {\bibfnamefont {B.~A.}\ \bibnamefont
  {Bernevig}},\ }\href@noop {} {\bibfield  {journal} {\bibinfo  {journal}
  {Physical Review X}\ }\textbf {\bibinfo {volume} {1}},\ \bibinfo {pages}
  {021014} (\bibinfo {year} {2011})}\BibitemShut {NoStop}%
\bibitem [{\citenamefont {Wu}\ \emph {et~al.}(2012)\citenamefont {Wu},
  \citenamefont {Bernevig},\ and\ \citenamefont {Regnault}}]{wu2012zoology}%
  \BibitemOpen
  \bibfield  {author} {\bibinfo {author} {\bibfnamefont {Y.-L.}\ \bibnamefont
  {Wu}}, \bibinfo {author} {\bibfnamefont {B.~A.}\ \bibnamefont {Bernevig}}, \
  and\ \bibinfo {author} {\bibfnamefont {N.}~\bibnamefont {Regnault}},\
  }\href@noop {} {\bibfield  {journal} {\bibinfo  {journal} {Physical Review
  B—Condensed Matter and Materials Physics}\ }\textbf {\bibinfo {volume}
  {85}},\ \bibinfo {pages} {075116} (\bibinfo {year} {2012})}\BibitemShut
  {NoStop}%
\bibitem [{\citenamefont {Bergholtz}\ and\ \citenamefont
  {Liu}(2013)}]{bergholtz2013topological}%
  \BibitemOpen
  \bibfield  {author} {\bibinfo {author} {\bibfnamefont {E.~J.}\ \bibnamefont
  {Bergholtz}}\ and\ \bibinfo {author} {\bibfnamefont {Z.}~\bibnamefont
  {Liu}},\ }\href@noop {} {\bibfield  {journal} {\bibinfo  {journal}
  {International Journal of Modern Physics B}\ }\textbf {\bibinfo {volume}
  {27}},\ \bibinfo {pages} {1330017} (\bibinfo {year} {2013})}\BibitemShut
  {NoStop}%
\bibitem [{\citenamefont {Grushin}\ \emph {et~al.}(2014)\citenamefont
  {Grushin}, \citenamefont {G{\'o}mez-Le{\'o}n},\ and\ \citenamefont
  {Neupert}}]{grushin2014floquet}%
  \BibitemOpen
  \bibfield  {author} {\bibinfo {author} {\bibfnamefont {A.~G.}\ \bibnamefont
  {Grushin}}, \bibinfo {author} {\bibfnamefont {{\'A}.}~\bibnamefont
  {G{\'o}mez-Le{\'o}n}}, \ and\ \bibinfo {author} {\bibfnamefont
  {T.}~\bibnamefont {Neupert}},\ }\href@noop {} {\bibfield  {journal} {\bibinfo
   {journal} {Physical review letters}\ }\textbf {\bibinfo {volume} {112}},\
  \bibinfo {pages} {156801} (\bibinfo {year} {2014})}\BibitemShut {NoStop}%
\bibitem [{\citenamefont {Wu}\ \emph {et~al.}(2013)\citenamefont {Wu},
  \citenamefont {Regnault},\ and\ \citenamefont {Bernevig}}]{wu2013bloch}%
  \BibitemOpen
  \bibfield  {author} {\bibinfo {author} {\bibfnamefont {Y.-L.}\ \bibnamefont
  {Wu}}, \bibinfo {author} {\bibfnamefont {N.}~\bibnamefont {Regnault}}, \ and\
  \bibinfo {author} {\bibfnamefont {B.~A.}\ \bibnamefont {Bernevig}},\
  }\href@noop {} {\bibfield  {journal} {\bibinfo  {journal} {Physical review
  letters}\ }\textbf {\bibinfo {volume} {110}},\ \bibinfo {pages} {106802}
  (\bibinfo {year} {2013})}\BibitemShut {NoStop}%
\bibitem [{\citenamefont {L{\"a}uchli}\ \emph {et~al.}(2013)\citenamefont
  {L{\"a}uchli}, \citenamefont {Liu}, \citenamefont {Bergholtz},\ and\
  \citenamefont {Moessner}}]{lauchli2013hierarchy}%
  \BibitemOpen
  \bibfield  {author} {\bibinfo {author} {\bibfnamefont {A.~M.}\ \bibnamefont
  {L{\"a}uchli}}, \bibinfo {author} {\bibfnamefont {Z.}~\bibnamefont {Liu}},
  \bibinfo {author} {\bibfnamefont {E.~J.}\ \bibnamefont {Bergholtz}}, \ and\
  \bibinfo {author} {\bibfnamefont {R.}~\bibnamefont {Moessner}},\ }\href@noop
  {} {\bibfield  {journal} {\bibinfo  {journal} {Physical Review Letters}\
  }\textbf {\bibinfo {volume} {111}},\ \bibinfo {pages} {126802} (\bibinfo
  {year} {2013})}\BibitemShut {NoStop}%
\bibitem [{\citenamefont {Neupert}\ \emph {et~al.}(2015)\citenamefont
  {Neupert}, \citenamefont {Chamon}, \citenamefont {Iadecola}, \citenamefont
  {Santos},\ and\ \citenamefont {Mudry}}]{neupert2015fractional}%
  \BibitemOpen
  \bibfield  {author} {\bibinfo {author} {\bibfnamefont {T.}~\bibnamefont
  {Neupert}}, \bibinfo {author} {\bibfnamefont {C.}~\bibnamefont {Chamon}},
  \bibinfo {author} {\bibfnamefont {T.}~\bibnamefont {Iadecola}}, \bibinfo
  {author} {\bibfnamefont {L.~H.}\ \bibnamefont {Santos}}, \ and\ \bibinfo
  {author} {\bibfnamefont {C.}~\bibnamefont {Mudry}},\ }\href@noop {}
  {\bibfield  {journal} {\bibinfo  {journal} {Physica Scripta}\ }\textbf
  {\bibinfo {volume} {2015}},\ \bibinfo {pages} {014005} (\bibinfo {year}
  {2015})}\BibitemShut {NoStop}%
\bibitem [{\citenamefont {Behrmann}\ \emph {et~al.}(2016)\citenamefont
  {Behrmann}, \citenamefont {Liu},\ and\ \citenamefont
  {Bergholtz}}]{behrmann2016model}%
  \BibitemOpen
  \bibfield  {author} {\bibinfo {author} {\bibfnamefont {J.}~\bibnamefont
  {Behrmann}}, \bibinfo {author} {\bibfnamefont {Z.}~\bibnamefont {Liu}}, \
  and\ \bibinfo {author} {\bibfnamefont {E.~J.}\ \bibnamefont {Bergholtz}},\
  }\href@noop {} {\bibfield  {journal} {\bibinfo  {journal} {Physical Review
  Letters}\ }\textbf {\bibinfo {volume} {116}},\ \bibinfo {pages} {216802}
  (\bibinfo {year} {2016})}\BibitemShut {NoStop}%
\bibitem [{\citenamefont {Yang}(2013)}]{yang2013geometry}%
  \BibitemOpen
  \bibfield  {author} {\bibinfo {author} {\bibfnamefont {K.}~\bibnamefont
  {Yang}},\ }\href@noop {} {\bibfield  {journal} {\bibinfo  {journal} {Physical
  Review B—Condensed Matter and Materials Physics}\ }\textbf {\bibinfo
  {volume} {88}},\ \bibinfo {pages} {241105} (\bibinfo {year}
  {2013})}\BibitemShut {NoStop}%
\bibitem [{\citenamefont {Zhu}\ \emph {et~al.}(2017)\citenamefont {Zhu},
  \citenamefont {Sodemann}, \citenamefont {Sheng},\ and\ \citenamefont
  {Fu}}]{zhu2017anisotropy}%
  \BibitemOpen
  \bibfield  {author} {\bibinfo {author} {\bibfnamefont {Z.}~\bibnamefont
  {Zhu}}, \bibinfo {author} {\bibfnamefont {I.}~\bibnamefont {Sodemann}},
  \bibinfo {author} {\bibfnamefont {D.}~\bibnamefont {Sheng}}, \ and\ \bibinfo
  {author} {\bibfnamefont {L.}~\bibnamefont {Fu}},\ }\href@noop {} {\bibfield
  {journal} {\bibinfo  {journal} {Physical Review B}\ }\textbf {\bibinfo
  {volume} {95}},\ \bibinfo {pages} {201116} (\bibinfo {year}
  {2017})}\BibitemShut {NoStop}%
\bibitem [{\citenamefont {He}\ \emph {et~al.}(2021)\citenamefont {He},
  \citenamefont {Yang}, \citenamefont {Goerbig},\ and\ \citenamefont
  {Mong}}]{he2021charge}%
  \BibitemOpen
  \bibfield  {author} {\bibinfo {author} {\bibfnamefont {Y.}~\bibnamefont
  {He}}, \bibinfo {author} {\bibfnamefont {K.}~\bibnamefont {Yang}}, \bibinfo
  {author} {\bibfnamefont {M.~O.}\ \bibnamefont {Goerbig}}, \ and\ \bibinfo
  {author} {\bibfnamefont {R.~S.}\ \bibnamefont {Mong}},\ }\href@noop {}
  {\bibfield  {journal} {\bibinfo  {journal} {Communications Physics}\ }\textbf
  {\bibinfo {volume} {4}},\ \bibinfo {pages} {116} (\bibinfo {year}
  {2021})}\BibitemShut {NoStop}%
\bibitem [{\citenamefont {Wang}\ \emph {et~al.}(2012)\citenamefont {Wang},
  \citenamefont {Narayanan}, \citenamefont {Wan},\ and\ \citenamefont
  {Zhang}}]{wang2012fractional}%
  \BibitemOpen
  \bibfield  {author} {\bibinfo {author} {\bibfnamefont {H.}~\bibnamefont
  {Wang}}, \bibinfo {author} {\bibfnamefont {R.}~\bibnamefont {Narayanan}},
  \bibinfo {author} {\bibfnamefont {X.}~\bibnamefont {Wan}}, \ and\ \bibinfo
  {author} {\bibfnamefont {F.}~\bibnamefont {Zhang}},\ }\href@noop {}
  {\bibfield  {journal} {\bibinfo  {journal} {Physical Review B—Condensed
  Matter and Materials Physics}\ }\textbf {\bibinfo {volume} {86}},\ \bibinfo
  {pages} {035122} (\bibinfo {year} {2012})}\BibitemShut {NoStop}%
\bibitem [{\citenamefont {Ippoliti}\ \emph {et~al.}(2017)\citenamefont
  {Ippoliti}, \citenamefont {Geraedts},\ and\ \citenamefont
  {Bhatt}}]{ippoliti2017numerical}%
  \BibitemOpen
  \bibfield  {author} {\bibinfo {author} {\bibfnamefont {M.}~\bibnamefont
  {Ippoliti}}, \bibinfo {author} {\bibfnamefont {S.~D.}\ \bibnamefont
  {Geraedts}}, \ and\ \bibinfo {author} {\bibfnamefont {R.~N.}\ \bibnamefont
  {Bhatt}},\ }\href@noop {} {\bibfield  {journal} {\bibinfo  {journal}
  {Physical Review B}\ }\textbf {\bibinfo {volume} {95}},\ \bibinfo {pages}
  {201104} (\bibinfo {year} {2017})}\BibitemShut {NoStop}%
\bibitem [{\citenamefont {Jin}\ \emph {et~al.}(2021)\citenamefont {Jin},
  \citenamefont {Tao}, \citenamefont {Li}, \citenamefont {Xu}, \citenamefont
  {Tang}, \citenamefont {Zhu}, \citenamefont {Liu}, \citenamefont {Watanabe},
  \citenamefont {Taniguchi}, \citenamefont {Hone} \emph
  {et~al.}}]{jin2021stripe}%
  \BibitemOpen
  \bibfield  {author} {\bibinfo {author} {\bibfnamefont {C.}~\bibnamefont
  {Jin}}, \bibinfo {author} {\bibfnamefont {Z.}~\bibnamefont {Tao}}, \bibinfo
  {author} {\bibfnamefont {T.}~\bibnamefont {Li}}, \bibinfo {author}
  {\bibfnamefont {Y.}~\bibnamefont {Xu}}, \bibinfo {author} {\bibfnamefont
  {Y.}~\bibnamefont {Tang}}, \bibinfo {author} {\bibfnamefont {J.}~\bibnamefont
  {Zhu}}, \bibinfo {author} {\bibfnamefont {S.}~\bibnamefont {Liu}}, \bibinfo
  {author} {\bibfnamefont {K.}~\bibnamefont {Watanabe}}, \bibinfo {author}
  {\bibfnamefont {T.}~\bibnamefont {Taniguchi}}, \bibinfo {author}
  {\bibfnamefont {J.~C.}\ \bibnamefont {Hone}},  \emph {et~al.},\ }\href@noop
  {} {\bibfield  {journal} {\bibinfo  {journal} {Nature Materials}\ }\textbf
  {\bibinfo {volume} {20}},\ \bibinfo {pages} {940} (\bibinfo {year}
  {2021})}\BibitemShut {NoStop}%
\bibitem [{\citenamefont {Carr}\ \emph {et~al.}(2017)\citenamefont {Carr},
  \citenamefont {Massatt}, \citenamefont {Fang}, \citenamefont {Cazeaux},
  \citenamefont {Luskin},\ and\ \citenamefont {Kaxiras}}]{carr2017twistronics}%
  \BibitemOpen
  \bibfield  {author} {\bibinfo {author} {\bibfnamefont {S.}~\bibnamefont
  {Carr}}, \bibinfo {author} {\bibfnamefont {D.}~\bibnamefont {Massatt}},
  \bibinfo {author} {\bibfnamefont {S.}~\bibnamefont {Fang}}, \bibinfo {author}
  {\bibfnamefont {P.}~\bibnamefont {Cazeaux}}, \bibinfo {author} {\bibfnamefont
  {M.}~\bibnamefont {Luskin}}, \ and\ \bibinfo {author} {\bibfnamefont
  {E.}~\bibnamefont {Kaxiras}},\ }\href@noop {} {\bibfield  {journal} {\bibinfo
   {journal} {Physical Review B}\ }\textbf {\bibinfo {volume} {95}},\ \bibinfo
  {pages} {075420} (\bibinfo {year} {2017})}\BibitemShut {NoStop}%
\bibitem [{\citenamefont {Wang}\ \emph
  {et~al.}(2021{\natexlab{a}})\citenamefont {Wang}, \citenamefont {Cui},
  \citenamefont {Jian}, \citenamefont {Cheng}, \citenamefont {Niu},
  \citenamefont {Yu}, \citenamefont {Yan},\ and\ \citenamefont
  {Huang}}]{wang2021stacking}%
  \BibitemOpen
  \bibfield  {author} {\bibinfo {author} {\bibfnamefont {S.}~\bibnamefont
  {Wang}}, \bibinfo {author} {\bibfnamefont {X.}~\bibnamefont {Cui}}, \bibinfo
  {author} {\bibfnamefont {C.}~\bibnamefont {Jian}}, \bibinfo {author}
  {\bibfnamefont {H.}~\bibnamefont {Cheng}}, \bibinfo {author} {\bibfnamefont
  {M.}~\bibnamefont {Niu}}, \bibinfo {author} {\bibfnamefont {J.}~\bibnamefont
  {Yu}}, \bibinfo {author} {\bibfnamefont {J.}~\bibnamefont {Yan}}, \ and\
  \bibinfo {author} {\bibfnamefont {W.}~\bibnamefont {Huang}},\ }\href@noop {}
  {\bibfield  {journal} {\bibinfo  {journal} {Advanced Materials}\ }\textbf
  {\bibinfo {volume} {33}},\ \bibinfo {pages} {2005735} (\bibinfo {year}
  {2021}{\natexlab{a}})}\BibitemShut {NoStop}%
\bibitem [{\citenamefont {Yu}\ \emph {et~al.}(2025)\citenamefont {Yu},
  \citenamefont {Yang}, \citenamefont {Kang}, \citenamefont {Huang},
  \citenamefont {Li},\ and\ \citenamefont {Wen}}]{yu2025engineering}%
  \BibitemOpen
  \bibfield  {author} {\bibinfo {author} {\bibfnamefont {F.}~\bibnamefont
  {Yu}}, \bibinfo {author} {\bibfnamefont {W.}~\bibnamefont {Yang}}, \bibinfo
  {author} {\bibfnamefont {J.}~\bibnamefont {Kang}}, \bibinfo {author}
  {\bibfnamefont {R.}~\bibnamefont {Huang}}, \bibinfo {author} {\bibfnamefont
  {L.}~\bibnamefont {Li}}, \ and\ \bibinfo {author} {\bibfnamefont
  {Y.}~\bibnamefont {Wen}},\ }\href@noop {} {\bibfield  {journal} {\bibinfo
  {journal} {Journal of Physics: Condensed Matter}\ }\textbf {\bibinfo {volume}
  {37}},\ \bibinfo {pages} {075502} (\bibinfo {year} {2025})}\BibitemShut
  {NoStop}%
\bibitem [{\citenamefont {Bi}\ \emph {et~al.}(2019)\citenamefont {Bi},
  \citenamefont {Yuan},\ and\ \citenamefont {Fu}}]{bi2019designing}%
  \BibitemOpen
  \bibfield  {author} {\bibinfo {author} {\bibfnamefont {Z.}~\bibnamefont
  {Bi}}, \bibinfo {author} {\bibfnamefont {N.~F.}\ \bibnamefont {Yuan}}, \ and\
  \bibinfo {author} {\bibfnamefont {L.}~\bibnamefont {Fu}},\ }\href@noop {}
  {\bibfield  {journal} {\bibinfo  {journal} {Physical Review B}\ }\textbf
  {\bibinfo {volume} {100}},\ \bibinfo {pages} {035448} (\bibinfo {year}
  {2019})}\BibitemShut {NoStop}%
\bibitem [{\citenamefont {Hu}\ \emph {et~al.}(2022)\citenamefont {Hu},
  \citenamefont {Zhang}, \citenamefont {Xie},\ and\ \citenamefont
  {Law}}]{hu2022nonlinear}%
  \BibitemOpen
  \bibfield  {author} {\bibinfo {author} {\bibfnamefont {J.-X.}\ \bibnamefont
  {Hu}}, \bibinfo {author} {\bibfnamefont {C.-P.}\ \bibnamefont {Zhang}},
  \bibinfo {author} {\bibfnamefont {Y.-M.}\ \bibnamefont {Xie}}, \ and\
  \bibinfo {author} {\bibfnamefont {K.}~\bibnamefont {Law}},\ }\href@noop {}
  {\bibfield  {journal} {\bibinfo  {journal} {Communications Physics}\ }\textbf
  {\bibinfo {volume} {5}},\ \bibinfo {pages} {255} (\bibinfo {year}
  {2022})}\BibitemShut {NoStop}%
\bibitem [{\citenamefont {Yang}\ \emph {et~al.}(2017)\citenamefont {Yang},
  \citenamefont {Hu}, \citenamefont {Lee},\ and\ \citenamefont
  {Papi{\'c}}}]{yang2017generalized}%
  \BibitemOpen
  \bibfield  {author} {\bibinfo {author} {\bibfnamefont {B.}~\bibnamefont
  {Yang}}, \bibinfo {author} {\bibfnamefont {Z.-X.}\ \bibnamefont {Hu}},
  \bibinfo {author} {\bibfnamefont {C.~H.}\ \bibnamefont {Lee}}, \ and\
  \bibinfo {author} {\bibfnamefont {Z.}~\bibnamefont {Papi{\'c}}},\ }\href@noop
  {} {\bibfield  {journal} {\bibinfo  {journal} {Physical review letters}\
  }\textbf {\bibinfo {volume} {118}},\ \bibinfo {pages} {146403} (\bibinfo
  {year} {2017})}\BibitemShut {NoStop}%
\bibitem [{\citenamefont {Mulligan}\ \emph {et~al.}(2010)\citenamefont
  {Mulligan}, \citenamefont {Nayak},\ and\ \citenamefont
  {Kachru}}]{mulligan2010isotropic}%
  \BibitemOpen
  \bibfield  {author} {\bibinfo {author} {\bibfnamefont {M.}~\bibnamefont
  {Mulligan}}, \bibinfo {author} {\bibfnamefont {C.}~\bibnamefont {Nayak}}, \
  and\ \bibinfo {author} {\bibfnamefont {S.}~\bibnamefont {Kachru}},\
  }\href@noop {} {\bibfield  {journal} {\bibinfo  {journal} {Physical Review
  B—Condensed Matter and Materials Physics}\ }\textbf {\bibinfo {volume}
  {82}},\ \bibinfo {pages} {085102} (\bibinfo {year} {2010})}\BibitemShut
  {NoStop}%
\bibitem [{\citenamefont {Yang}\ \emph {et~al.}(2012)\citenamefont {Yang},
  \citenamefont {Papi{\'c}}, \citenamefont {Rezayi}, \citenamefont {Bhatt},\
  and\ \citenamefont {Haldane}}]{yang2012band}%
  \BibitemOpen
  \bibfield  {author} {\bibinfo {author} {\bibfnamefont {B.}~\bibnamefont
  {Yang}}, \bibinfo {author} {\bibfnamefont {Z.}~\bibnamefont {Papi{\'c}}},
  \bibinfo {author} {\bibfnamefont {E.}~\bibnamefont {Rezayi}}, \bibinfo
  {author} {\bibfnamefont {R.}~\bibnamefont {Bhatt}}, \ and\ \bibinfo {author}
  {\bibfnamefont {F.}~\bibnamefont {Haldane}},\ }\href@noop {} {\bibfield
  {journal} {\bibinfo  {journal} {Physical Review B—Condensed Matter and
  Materials Physics}\ }\textbf {\bibinfo {volume} {85}},\ \bibinfo {pages}
  {165318} (\bibinfo {year} {2012})}\BibitemShut {NoStop}%
\bibitem [{\citenamefont {Tarnopolsky}\ \emph {et~al.}(2019)\citenamefont
  {Tarnopolsky}, \citenamefont {Kruchkov},\ and\ \citenamefont
  {Vishwanath}}]{tarnopolsky2019origin}%
  \BibitemOpen
  \bibfield  {author} {\bibinfo {author} {\bibfnamefont {G.}~\bibnamefont
  {Tarnopolsky}}, \bibinfo {author} {\bibfnamefont {A.~J.}\ \bibnamefont
  {Kruchkov}}, \ and\ \bibinfo {author} {\bibfnamefont {A.}~\bibnamefont
  {Vishwanath}},\ }\href@noop {} {\bibfield  {journal} {\bibinfo  {journal}
  {Physical review letters}\ }\textbf {\bibinfo {volume} {122}},\ \bibinfo
  {pages} {106405} (\bibinfo {year} {2019})}\BibitemShut {NoStop}%
\bibitem [{\citenamefont {Wang}\ \emph
  {et~al.}(2021{\natexlab{b}})\citenamefont {Wang}, \citenamefont {Zheng},
  \citenamefont {Millis},\ and\ \citenamefont {Cano}}]{wang2021chiral}%
  \BibitemOpen
  \bibfield  {author} {\bibinfo {author} {\bibfnamefont {J.}~\bibnamefont
  {Wang}}, \bibinfo {author} {\bibfnamefont {Y.}~\bibnamefont {Zheng}},
  \bibinfo {author} {\bibfnamefont {A.~J.}\ \bibnamefont {Millis}}, \ and\
  \bibinfo {author} {\bibfnamefont {J.}~\bibnamefont {Cano}},\ }\href@noop {}
  {\bibfield  {journal} {\bibinfo  {journal} {Physical Review Research}\
  }\textbf {\bibinfo {volume} {3}},\ \bibinfo {pages} {023155} (\bibinfo {year}
  {2021}{\natexlab{b}})}\BibitemShut {NoStop}%
\bibitem [{\citenamefont {Wang}\ \emph
  {et~al.}(2021{\natexlab{c}})\citenamefont {Wang}, \citenamefont {Cano},
  \citenamefont {Millis}, \citenamefont {Liu},\ and\ \citenamefont
  {Yang}}]{wang2021exact}%
  \BibitemOpen
  \bibfield  {author} {\bibinfo {author} {\bibfnamefont {J.}~\bibnamefont
  {Wang}}, \bibinfo {author} {\bibfnamefont {J.}~\bibnamefont {Cano}}, \bibinfo
  {author} {\bibfnamefont {A.~J.}\ \bibnamefont {Millis}}, \bibinfo {author}
  {\bibfnamefont {Z.}~\bibnamefont {Liu}}, \ and\ \bibinfo {author}
  {\bibfnamefont {B.}~\bibnamefont {Yang}},\ }\href@noop {} {\bibfield
  {journal} {\bibinfo  {journal} {Physical review letters}\ }\textbf {\bibinfo
  {volume} {127}},\ \bibinfo {pages} {246403} (\bibinfo {year}
  {2021}{\natexlab{c}})}\BibitemShut {NoStop}%
\bibitem [{\citenamefont {Ferrari}(1990)}]{ferrari1990two}%
  \BibitemOpen
  \bibfield  {author} {\bibinfo {author} {\bibfnamefont {R.}~\bibnamefont
  {Ferrari}},\ }\href@noop {} {\bibfield  {journal} {\bibinfo  {journal}
  {Physical Review B}\ }\textbf {\bibinfo {volume} {42}},\ \bibinfo {pages}
  {4598} (\bibinfo {year} {1990})}\BibitemShut {NoStop}%
\bibitem [{\citenamefont {Ferrari}(1995)}]{ferrari1995wannier}%
  \BibitemOpen
  \bibfield  {author} {\bibinfo {author} {\bibfnamefont {R.}~\bibnamefont
  {Ferrari}},\ }\href@noop {} {\bibfield  {journal} {\bibinfo  {journal}
  {International Journal of Modern Physics B}\ }\textbf {\bibinfo {volume}
  {9}},\ \bibinfo {pages} {3333} (\bibinfo {year} {1995})}\BibitemShut
  {NoStop}%
\bibitem [{\citenamefont {Haldane}(2018)}]{haldane2018modular}%
  \BibitemOpen
  \bibfield  {author} {\bibinfo {author} {\bibfnamefont {F.}~\bibnamefont
  {Haldane}},\ }\href@noop {} {\bibfield  {journal} {\bibinfo  {journal}
  {Journal of Mathematical Physics}\ }\textbf {\bibinfo {volume} {59}}
  (\bibinfo {year} {2018})}\BibitemShut {NoStop}%
\bibitem [{\citenamefont {Wang}\ \emph {et~al.}(2019)\citenamefont {Wang},
  \citenamefont {Geraedts}, \citenamefont {Rezayi},\ and\ \citenamefont
  {Haldane}}]{wang2019lattice}%
  \BibitemOpen
  \bibfield  {author} {\bibinfo {author} {\bibfnamefont {J.}~\bibnamefont
  {Wang}}, \bibinfo {author} {\bibfnamefont {S.~D.}\ \bibnamefont {Geraedts}},
  \bibinfo {author} {\bibfnamefont {E.}~\bibnamefont {Rezayi}}, \ and\ \bibinfo
  {author} {\bibfnamefont {F.}~\bibnamefont {Haldane}},\ }\href@noop {}
  {\bibfield  {journal} {\bibinfo  {journal} {Physical Review B}\ }\textbf
  {\bibinfo {volume} {99}},\ \bibinfo {pages} {125123} (\bibinfo {year}
  {2019})}\BibitemShut {NoStop}%
\bibitem [{sup()}]{supple}%
  \BibitemOpen
  \href@noop {} {}\bibinfo {note} {The Supplemental Material contains
  additional theoretical and numerical results.}\BibitemShut {Stop}%
\bibitem [{\citenamefont {Qiu}\ \emph {et~al.}(2012)\citenamefont {Qiu},
  \citenamefont {Haldane}, \citenamefont {Wan}, \citenamefont {Yang},\ and\
  \citenamefont {Yi}}]{Qiu2012modelanisotropicH}%
  \BibitemOpen
  \bibfield  {author} {\bibinfo {author} {\bibfnamefont {R.-Z.}\ \bibnamefont
  {Qiu}}, \bibinfo {author} {\bibfnamefont {F.~D.~M.}\ \bibnamefont {Haldane}},
  \bibinfo {author} {\bibfnamefont {X.}~\bibnamefont {Wan}}, \bibinfo {author}
  {\bibfnamefont {K.}~\bibnamefont {Yang}}, \ and\ \bibinfo {author}
  {\bibfnamefont {S.}~\bibnamefont {Yi}},\ }\href@noop {} {\bibfield  {journal}
  {\bibinfo  {journal} {Physical Review B}\ }\textbf {\bibinfo {volume} {85}},\
  \bibinfo {pages} {115308} (\bibinfo {year} {2012})}\BibitemShut {NoStop}%
\bibitem [{\citenamefont {Haldane}(2011)}]{Haldane2011anisotropygeometry}%
  \BibitemOpen
  \bibfield  {author} {\bibinfo {author} {\bibfnamefont {F.~D.~M.}\
  \bibnamefont {Haldane}},\ }\href@noop {} {\bibfield  {journal} {\bibinfo
  {journal} {Physical Review Letters}\ }\textbf {\bibinfo {volume} {107}},\
  \bibinfo {pages} {116801} (\bibinfo {year} {2011})}\BibitemShut {NoStop}%
\bibitem [{\citenamefont {You}\ \emph {et~al.}(2014)\citenamefont {You},
  \citenamefont {Cho},\ and\ \citenamefont {Fradkin}}]{you2014theory}%
  \BibitemOpen
  \bibfield  {author} {\bibinfo {author} {\bibfnamefont {Y.}~\bibnamefont
  {You}}, \bibinfo {author} {\bibfnamefont {G.~Y.}\ \bibnamefont {Cho}}, \ and\
  \bibinfo {author} {\bibfnamefont {E.}~\bibnamefont {Fradkin}},\ }\href@noop
  {} {\bibfield  {journal} {\bibinfo  {journal} {Physical Review X}\ }\textbf
  {\bibinfo {volume} {4}},\ \bibinfo {pages} {041050} (\bibinfo {year}
  {2014})}\BibitemShut {NoStop}%
\bibitem [{\citenamefont {Shibata}\ and\ \citenamefont
  {Yoshioka}(2001)}]{shibata2001ground}%
  \BibitemOpen
  \bibfield  {author} {\bibinfo {author} {\bibfnamefont {N.}~\bibnamefont
  {Shibata}}\ and\ \bibinfo {author} {\bibfnamefont {D.}~\bibnamefont
  {Yoshioka}},\ }\href@noop {} {\bibfield  {journal} {\bibinfo  {journal}
  {Physical Review Letters}\ }\textbf {\bibinfo {volume} {86}},\ \bibinfo
  {pages} {5755} (\bibinfo {year} {2001})}\BibitemShut {NoStop}%
\bibitem [{\citenamefont {Yang}\ \emph {et~al.}(2023)\citenamefont {Yang},
  \citenamefont {Bai}, \citenamefont {Zibrov}, \citenamefont {Joy},
  \citenamefont {Taniguchi}, \citenamefont {Watanabe}, \citenamefont {Skinner},
  \citenamefont {Goerbig},\ and\ \citenamefont {Young}}]{yang2023cascade}%
  \BibitemOpen
  \bibfield  {author} {\bibinfo {author} {\bibfnamefont {F.}~\bibnamefont
  {Yang}}, \bibinfo {author} {\bibfnamefont {R.}~\bibnamefont {Bai}}, \bibinfo
  {author} {\bibfnamefont {A.~A.}\ \bibnamefont {Zibrov}}, \bibinfo {author}
  {\bibfnamefont {S.}~\bibnamefont {Joy}}, \bibinfo {author} {\bibfnamefont
  {T.}~\bibnamefont {Taniguchi}}, \bibinfo {author} {\bibfnamefont
  {K.}~\bibnamefont {Watanabe}}, \bibinfo {author} {\bibfnamefont
  {B.}~\bibnamefont {Skinner}}, \bibinfo {author} {\bibfnamefont {M.~O.}\
  \bibnamefont {Goerbig}}, \ and\ \bibinfo {author} {\bibfnamefont {A.~F.}\
  \bibnamefont {Young}},\ }\href@noop {} {\bibfield  {journal} {\bibinfo
  {journal} {Physical Review Letters}\ }\textbf {\bibinfo {volume} {131}},\
  \bibinfo {pages} {226501} (\bibinfo {year} {2023})}\BibitemShut {NoStop}%
\bibitem [{\citenamefont {Rezayi}\ and\ \citenamefont
  {Haldane}(2000)}]{Rezayi2000Coulomb}%
  \BibitemOpen
  \bibfield  {author} {\bibinfo {author} {\bibfnamefont {E.~H.}\ \bibnamefont
  {Rezayi}}\ and\ \bibinfo {author} {\bibfnamefont {F.~D.~M.}\ \bibnamefont
  {Haldane}},\ }\href@noop {} {\bibfield  {journal} {\bibinfo  {journal}
  {Physical review letters}\ }\textbf {\bibinfo {volume} {84}},\ \bibinfo
  {pages} {4685} (\bibinfo {year} {2000})}\BibitemShut {NoStop}%
\bibitem [{\citenamefont {Bergholtz}\ \emph {et~al.}(2006)\citenamefont
  {Bergholtz}, \citenamefont {Kailasvuori}, \citenamefont {Wikberg},
  \citenamefont {Hansson},\ and\ \citenamefont
  {Karlhede}}]{Bergholtz2006threefold}%
  \BibitemOpen
  \bibfield  {author} {\bibinfo {author} {\bibfnamefont {E.~J.}\ \bibnamefont
  {Bergholtz}}, \bibinfo {author} {\bibfnamefont {J.}~\bibnamefont
  {Kailasvuori}}, \bibinfo {author} {\bibfnamefont {E.}~\bibnamefont
  {Wikberg}}, \bibinfo {author} {\bibfnamefont {T.~H.}\ \bibnamefont
  {Hansson}}, \ and\ \bibinfo {author} {\bibfnamefont {A.}~\bibnamefont
  {Karlhede}},\ }\href@noop {} {\bibfield  {journal} {\bibinfo  {journal}
  {Physical Review B—Condensed Matter and Materials Physics}\ }\textbf
  {\bibinfo {volume} {74}},\ \bibinfo {pages} {081308} (\bibinfo {year}
  {2006})}\BibitemShut {NoStop}%
\bibitem [{\citenamefont {Huang}\ \emph {et~al.}(2023)\citenamefont {Huang},
  \citenamefont {Wu}, \citenamefont {Hu}, \citenamefont {Cai}, \citenamefont
  {Li}, \citenamefont {An}, \citenamefont {Feng}, \citenamefont {Ye},
  \citenamefont {Lin}, \citenamefont {Law} \emph {et~al.}}]{huang2023giant}%
  \BibitemOpen
  \bibfield  {author} {\bibinfo {author} {\bibfnamefont {M.}~\bibnamefont
  {Huang}}, \bibinfo {author} {\bibfnamefont {Z.}~\bibnamefont {Wu}}, \bibinfo
  {author} {\bibfnamefont {J.}~\bibnamefont {Hu}}, \bibinfo {author}
  {\bibfnamefont {X.}~\bibnamefont {Cai}}, \bibinfo {author} {\bibfnamefont
  {E.}~\bibnamefont {Li}}, \bibinfo {author} {\bibfnamefont {L.}~\bibnamefont
  {An}}, \bibinfo {author} {\bibfnamefont {X.}~\bibnamefont {Feng}}, \bibinfo
  {author} {\bibfnamefont {Z.}~\bibnamefont {Ye}}, \bibinfo {author}
  {\bibfnamefont {N.}~\bibnamefont {Lin}}, \bibinfo {author} {\bibfnamefont
  {K.~T.}\ \bibnamefont {Law}},  \emph {et~al.},\ }\href@noop {} {\bibfield
  {journal} {\bibinfo  {journal} {National Science Review}\ }\textbf {\bibinfo
  {volume} {10}},\ \bibinfo {pages} {nwac232} (\bibinfo {year}
  {2023})}\BibitemShut {NoStop}%
\bibitem [{\citenamefont {Zhang}\ \emph {et~al.}(2022)\citenamefont {Zhang},
  \citenamefont {Xiao}, \citenamefont {Zhou}, \citenamefont {Hu}, \citenamefont
  {Xie}, \citenamefont {Yan},\ and\ \citenamefont {Law}}]{zhang2022giant}%
  \BibitemOpen
  \bibfield  {author} {\bibinfo {author} {\bibfnamefont {C.-P.}\ \bibnamefont
  {Zhang}}, \bibinfo {author} {\bibfnamefont {J.}~\bibnamefont {Xiao}},
  \bibinfo {author} {\bibfnamefont {B.~T.}\ \bibnamefont {Zhou}}, \bibinfo
  {author} {\bibfnamefont {J.-X.}\ \bibnamefont {Hu}}, \bibinfo {author}
  {\bibfnamefont {Y.-M.}\ \bibnamefont {Xie}}, \bibinfo {author} {\bibfnamefont
  {B.}~\bibnamefont {Yan}}, \ and\ \bibinfo {author} {\bibfnamefont {K.~T.}\
  \bibnamefont {Law}},\ }\href@noop {} {\bibfield  {journal} {\bibinfo
  {journal} {Physical Review B}\ }\textbf {\bibinfo {volume} {106}},\ \bibinfo
  {pages} {L041111} (\bibinfo {year} {2022})}\BibitemShut {NoStop}%
\bibitem [{\citenamefont {Zhou}\ \emph {et~al.}(2022)\citenamefont {Zhou},
  \citenamefont {Egan},\ and\ \citenamefont {Franz}}]{zhou2022moire}%
  \BibitemOpen
  \bibfield  {author} {\bibinfo {author} {\bibfnamefont {B.~T.}\ \bibnamefont
  {Zhou}}, \bibinfo {author} {\bibfnamefont {S.}~\bibnamefont {Egan}}, \ and\
  \bibinfo {author} {\bibfnamefont {M.}~\bibnamefont {Franz}},\ }\href@noop {}
  {\bibfield  {journal} {\bibinfo  {journal} {Physical Review Research}\
  }\textbf {\bibinfo {volume} {4}},\ \bibinfo {pages} {L012032} (\bibinfo
  {year} {2022})}\BibitemShut {NoStop}%
\end{thebibliography}
%

\end{document}